\documentclass[12pt]{article}

\usepackage[table]{xcolor}
\usepackage[normalem]{ulem}
\usepackage{longtable}
\usepackage{footmisc}

\usepackage{scicite}
\usepackage{ref_macros_sci}

\usepackage{soul}

\usepackage{comment}

\usepackage{cancel}
\usepackage[normalem]{ulem}
\usepackage{graphicx}
\usepackage{setspace}

\usepackage{amsmath}
\usepackage{amssymb}
\usepackage{xspace}
\usepackage{multirow}
\usepackage{array}
\newcolumntype{C}[1]{>{\centering\arraybackslash}p{#1}}
\usepackage{rotating}

\usepackage{siunitx}

\def\bibcommenthead{}

\DeclareRobustCommand{\ion}[2]{%
\relax\ifmmode
\ifx\testbx\f@series
{\mathbf{#1\,\mathsc{#2}}}\else
{\mathrm{#1\,\mathsc{#2}}}\fi
\else\textup{#1\,{\mdseries\textsc{#2}}}%
\fi}

\newcommand{\msun}{$M_{\odot}$\xspace}

\newcommand\nicer{\textit{NICER}\xspace}

\newcommand{\xmm}{{\it XMM-Newton}\xspace}
\newcommand{\targetlong}{Swift J023017.0+283603\xspace}
\newcommand{\target}{Swift\,J0230+28\xspace}
\newcommand{\swift}{\textit{Swift}\xspace}
\newcommand\swiftall{\textit{Neil Gehrels Swift Observatory}\xspace}
\newcommand{\ergs}{{erg~s$^{-1}$}\xspace}
\newcommand{\fluxunits}{{erg~s$^{-1}$~cm$^{-2}$}\xspace}
\newcommand\arcsec{\mbox{$^{\prime\prime}$}\xspace}
\newcommand\arcmin{\mbox{$^{\prime}$}\xspace}

\newcommand{\mbh}{$M_{\rm BH}$\xspace}
\newcommand{\msol}{$M_\odot$\xspace}
\newcommand{\oiii}{[\ion{O}{iii}]\xspace}
\newcommand{\Lbol}{\ensuremath{L_{bol}}\xspace}
\newcommand\srge{\textit{SRG}/eROSITA\xspace}
\newcommand{\Lbolquie}{\ensuremath{L_{\rm quiet,bol}}\xspace}

\newcommand{\LEdd}{\ensuremath{L_{\rm Edd}}\xspace}

\renewcommand{\figurename}{Figure}
\renewcommand{\tablename}{Table}

\usepackage[final]{pdfpages}

\usepackage{times}
\usepackage[hypcap=false]{caption} 
\usepackage{subcaption, booktabs}

\usepackage{floatrow}
\DeclareFloatVCode{somespace}{\vspace{1.667\baselineskip}}
\usepackage{hyperref}

\newenvironment{sciabstract}{%
\begin{quote} \bf}
{\end{quote}}

\newcounter{lastnote}

\title{X-ray eruptions every 22 days from the nucleus of a nearby galaxy} 

\author{Muryel Guolo$^{1}$,
Dheeraj R. Pasham$^{2}$,
Michal Zajaček$^{3}$,
Eric R.~Coughlin$^{4}$,\\
Suvi Gezari$^{5,1}$,
Petra Sukov\'a$^{6}$,
Thomas Wevers$^{5,7}$,
Vojt{\v e}ch Witzany$^{8}$,\\
Francesco Tombesi$^{9,10,11,12,13}$,
Sjoert van Velzen$^{14}$,
Kate D. Alexander$^{15}$, \\
Yuhan Yao$^{16,17}$,
Riccardo Arcodia$^{2}$,
Vladim\'{\i}r Karas$^{6}$,
James C.~A. Miller-Jones$^{18}$,\\
Ronald Remillard$^{2}$,
Keith Gendreau$^{13}$,
Elizabeth~C.~Ferrara$^{12,19,13}$\\
{\small $^{\bf 1}$Department of Physics and Astronomy, Johns Hopkins University, MD, USA}\\
{\small $^{\bf 2}$Kavli Institute for Astrophysics and Space Research, Massachusetts Institute of Technology, Cambridge, MA, USA}\\
{\small $^{\bf 3}$Department of Theoretical physics and Astrophysics, Masaryk University, Czech Republic}\\
{\small $^{\bf 4}$Department of Physics, Syracuse University, Syracuse, NY, USA}\\
{\small $^{\bf 5}$Space Telescope Science Institute, Baltimore, MD, US}\\
{\small $^{\bf 6}$Astronomical Institute of the Czech Academy of Sciences, Prague, Czech Republic}\\
{\small $^{\bf 7}$European Southern Observatory, Chile}\\
{\small $^{\bf 8}$Institute of Theoretical Physics, Charles University, Prague, Czech Republic}\\
{\small $^{\bf 9}$Physics Department, Tor Vergata University of Rome,  Rome, Italy }\\
{\small $^{\bf 10}$INAF – Astronomical Observatory of Rome, Rome, Italy  }\\
{\small $^{\bf 11}$INFN - Roma Tor Vergata, Rome, Italy }\\
{\small $^{\bf 12}$Department of Astronomy, University of Maryland, MD, USA}\\
{\small $^{\bf 13}$NASA Goddard Space Flight Center, Code 662, Greenbelt, MD, USA }\\
{\small $^{\bf 14}$Leiden Observatory, Leiden University, The Netherlands }\\
{\small $^{\bf 15}$Steward Observatory, University of Arizona, AZ, USA}\\
{\small $^{\bf 16}$Miller Institute for Basic Research in Science, 468 Donner Lab, Berkeley, CA 94720, USA}\\
{\small $^{\bf 17}$Department of Astronomy, University of California, Berkeley, CA 94720, USA}\\
{\small $^{\bf 18}$International Centre for Radio Astronomy Research - Curtin University, WA, Australia}\\\
{\small $^{\bf 19}$Center for Exploration and Space Studies (CRESST), NASA/GSFC, MD, USA}\\ \and
\normalsize{$^\ast$To whom correspondence should be addressed; E-mail:  mguolop1@jhu.edu}
}
\date{}

\begin{document}

\baselineskip24pt

\maketitle 

\newpage
\begin{sciabstract}

Galactic nuclei showing recurrent phases of activity and quiescence have recently been discovered, with recurrence times as short as a few hours to a day -- known as quasi-periodic X-ray eruption (QPE) sources
-- to as long as hundreds to a thousand days for repeating nuclear transients (RNTs). Here we present a multi-wavelength overview of \targetlong (hereafter \target), a source that exhibits repeating and quasi-periodic X-ray flares from the nucleus of a previously unremarkable galaxy at $\sim$ 165 Mpc, with a recurrence time of approximately 22 days, an intermediary timescale between known RNTs and QPE sources.  
The source also shows transient radio emission, likely associated with the X-ray emission. Such recurrent soft X-ray eruptions, with no accompanying UV/optical emission, are strikingly similar to QPE sources.
However, in addition to having a recurrence time that is $\sim 25$ times longer than the longest-known QPE source, \target's eruptions exhibit somewhat distinct shapes and temperature evolution than the known QPE sources. Scenarios involving extreme mass ratio inspirals are favored over disk instability models. The source reveals an unexplored timescale for repeating extragalactic transients and highlights the need for a wide-field, time-domain X-ray mission to explore the parameter space of recurring X-ray transients.

--–– 
\end{sciabstract}

The field of view of \targetlong (hereafter \target) was first observed by \swiftall X-Ray Telescope (XRT)  between December 2021 and January 
2022 following the discovery of the supernova SN2021afkk\cite{SN2021afkk} -- located $\sim$ 4\arcmin from \target~-- during which no X-ray emission was detected from the 
position of \target (stacked upper-limit 0.3-2.0 keV flux of $2\times10^{-14}$ \fluxunits). However, an observation taken on 22 
June 2022, processed by \swift live catalog of transients \cite{xrttrans,Evans2023}, revealed an X-ray source with a 0.3-2.0 keV flux of $7\times10^{-13}$ \fluxunits,
suggesting an enhancement of more than a factor of 35 \cite{css_atel}. 
The X-ray spectrum was soft and thermal with a temperature of 
$121^{+13}_{-25}$ eV ($1.4^{+0.2}_{-0.3}\times 10^6\,{\rm K}$).
Based on the spatial coincidence of the 
source with the center of a nearby galaxy, the 
soft/thermal X-ray spectrum, and the lack of any previous X-ray detection -- see ``Constraints on the start of the eruptions'' and Extended Data Figure (EDF) 1 in Methods -- it was initially reported \cite{css_atel,css_atel2} to be a 
flare resulting from the tidal disruption of a star by a 
massive black hole (MBH). 

Monitoring 
of \target with \swift/XRT
between June and August of 2022 (MJD 59752-59798) revealed X-ray eruptions (increases in the 0.3--2.0 keV X-ray flux from non-detection to peaks of a factor of up to 100 higher than the upper limits) that lasted several days and were separated by longer periods ($\gtrsim 15$ days) of non-detections (see top panel of Fig. 1). The X-ray eruptions had no accompanying changes in the optical and the UV bands (upper-limit in the host-subtracted UV luminosity $\lesssim 3 \times 10^{42}$ \ergs, see ``UV/optical and radio counterparts'', in Methods). 
These properties exclude the interpretation of a \emph{classical} tidal disruption event (TDE, (e.g., \cite{Rees1988, van_Velzen_21, Hammerstein2023,Guolo2023}). 
Following the indication of these recurrent X-ray eruptions, we initiated a high-cadence monitoring program with the Neutron Star Interior Composition Explorer (\nicer) starting on MJD 59798 as well as multi-wavelength follow-up with several other facilities (see ``Observations and Data Analysis'' in Methods).

In the eight months following \target's discovery, \swift/XRT observed the source with
$\sim$ daily cadence for a total of $\sim 160$ ks, while \nicer observed the source multiple 
times per day -- albeit with some gaps -- for a total of $\sim 500$ ks. The monitoring campaign 
confirmed the recurrence of several X-ray eruptions, that were apparently repeating 
every $\sim$ 3 weeks, as shown in Fig.~1. The Lomb-Scargle periodogram 
(LSP, \cite{scargle,lspnorm}) of the resulting light curve, as shown in Fig.~2, shows a strong peak 
at $21.8^{+1.2}_{-0.5}$ days, which indicates that the 
eruptions are quasi-periodic in nature. The pink shaded 
regions in Fig.~1 mark time intervals separated 
by the $21.8^{+1.2}_{-0.5}$ day period, which are referred to as epochs (\textit{E}), from \textit{E1}, \textit{E2}, \ldots, \textit{E11}, during the $\sim 240$ days of monitoring. 

Around most of the marked epochs, namely \textit{E1}-\textit{E6} and \textit{E10}-\textit{E11}, \target showed high amplitude eruptions with a mean full-width at half maximum duration of $\sim$ 4.5 days (see Supplementary Information Table 1). However, around \textit{E7} and \textit{E8} instead of the few-days-long eruptions, short-lived ($< 1$ day) and lower-amplitude eruptions were observed. Furthermore, no X-ray detections were observed around \textit{E9}, though extremely short-lived eruptions cannot be excluded given the lack of high cadence \nicer observations at the time. This indicates that, although the eruptions in \target are quasi-periodic, a certain degree of irregularity in the system's periodicity and amplitude is present. The eruptions in \target are slightly asymmetric; an asymmetric Gaussian profile fitting to the shape of the well-sampled and days-long eruptions shows that the rises are $\sim 30\%$ longer than the decays, see ``X-ray light curve'', in Methods, and EDF 2.

The high count rate obtained by \nicer allows us to perform time-resolved spectral analyses during the eruptions (see ``Time-resolved X-ray analyses'' in Methods). The X-ray spectra are soft with no photons detected at energies greater than 1.5 keV over all eruptions, and a thermal model modified by absorption --  both Galactic and intrinsic  (column density $N_{\rm H} \approx 1-3 \times 10^{20} \ \rm{cm}^{-2}$) -- fits the spectra reasonably well -- $\chi^2$/degrees of freedom $\in (0.9,1.9)$ -- as shown by Fig.~3 and EDF 3. The best-fitted temperature varies (see Fig.~4, and Extended Data Table 1) with time, and has a mean (standard deviation) value of 160 eV (50 eV). The variations in temperature correlate with the evolution of the eruptions: a simple regression between temperature and 0.3-0.8 keV X-ray luminosity shows a correlation in the form $L \propto T^{1.9\pm0.5}$. However, it should be noted that the large spread is driven by the fact that \target does not show a cool $\rightarrow$ warm $\rightarrow$ cool temperature evolution in each eruption. The temperature increases from the rise ($\sim$ 100 eV) to the peak ($\sim$ 150 eV) of the eruptions, but instead of decreasing during the decay -- as one would expect for a direct correlation between temperature and luminosity -- it continuous to increase, up to $\sim$ 200 eV, hence it shows a cool $\rightarrow$ warm temperature evolution.

Our radio monitoring (see ``Very Large Array (VLA)'' and EDF 4, in Methods) shows a transient point-like source at \target's position. While the first and third observations (MJD 59762 and MJD 59933) showed no detections with 3$\sigma$ upper limits of 15 and 25 $\mu$Jy, respectively, our second visit on MJD 59842 shows a detection with a flux of 93$\pm$7 $\mu$Jy. This radio detection coincides with one of the X-ray eruptions, while the two radio non-detections coincide with X-ray quiescent phases suggesting that the X-ray eruptions may be accompanied by radio emission (Fig.~1).

The position derived from XRT for \target is consistent  with the nucleus (0.2\arcsec $\pm$ 3.6\arcsec, 90\% uncertainty, from the photometric center) of a spiral galaxy (see EDF 5) at 165 Mpc (z = 0.036). The host galaxy is not in any active galactic nuclei (AGN) catalog; there is no archival detection in the X-rays, nor in the radio, prior to the start of the eruptions. The host also shows no infrared (IR) photometric excess, nor any variability in the IR bands, that would indicate the presence of a hot dust component (i.e., a ``torus''). However, optical emission line diagnostic diagrams (see EDF 6) based on nuclear spectra indicate the presence of a weak AGN (or low-luminosity AGN, LLAGN). The high-resolution optical spectrum of the nuclear region also allowed us to measure the stellar population velocity 
dispersion ($\sigma_*$) and estimate a black hole mass (\mbh) of log (\mbh/\msun) = $6.6 \pm 0.4$, using the standard \mbh-$\sigma_*$ relation \cite{Gultekin2009}, which is in agreement with the \mbh derived from the host galaxy mass ($M_*$, see EDF 7) assuming the \mbh-$M_*$ relation of \cite{Greene2020} .
Based on the \oiii emission line and the 2 -- 10 keV luminosity upper limit, we estimate the upper limit for the bolometric luminosity of the AGN (prior to the start of eruptions) to be \Lbolquie $\leq 9 \times 10^{41}$ \ergs, which for the 
estimated black hole mass translates into an Eddington ratio ($\lambda_{\rm Edd} = \Lbolquie/L_{\rm Edd}$) $<$ 0.002, supporting the LLAGN classification by line ratio diagnostic diagrams (see ``The host galaxy'' in Methods). At such a low accretion rate, a standard thin accretion disc should not be present in \target's host, instead, any accretion flow, if present, is more likely an Advection Dominated Accretion Flow (ADAF).

The recurrent phases of high activity followed by phases of quiescence, could, in principle, classify \target as a repeating nuclear transient (RNTs). Three clear cases of RNTs are known: ASASSN-14ko \cite{Payne2021}, eRASStJ045650.3-203750  (hereafter eRA J0456-20, \cite{Liu2023}) and AT2018fyk \cite{Wevers2023}. These sources show repeated flares, with recurrence times varying from 114-1200 days (see ``Repeating Nuclear Transients and \target'', Supplementary Information, for detailed properties of these sources), and are interpreted by most studies as the result of a star being repeatedly partially disrupted by the central MBH \cite{Payne2021,Cufari2022,Wevers2023,Liu2023,Liu2023b}. However, \target differs significantly from RNTs in the following ways: 
i) RNTs exhibit much brighter X-ray luminosities compared to \target; ii) RNTs are even brighter in the UV/optical bands than in X-rays, while \target does not exhibit any detected UV/optical emission; 
with rapid rises and much slower decays, while \target's eruptions have a slightly slower rise than decay; iii) none of the RNTs show a purely soft/thermal X-ray spectrum, like in \target.

\target's recurrent soft X-ray eruptions with no accompanying UV/optical emission from the nucleus of a galaxy hosting a relatively small black hole (\mbh $\leq 10^{6.6}$ \msol), are characteristics of the recently discovered class of quasi-periodic X-ray eruption (QPE) sources. Four confirmed QPE sources are known: GSN 069 \cite{Miniutti2019}, RX J1301 \cite{Giustini2020}, eRO-QPE1 and eRO-QPE2 \cite{Arcodia2021}, and a detailed description of their properties are presented in ``Quasi-periodic erupters and Swift J0230+28'' in Methods. However, the known QPE sources show much shorter mean recurrence times than \target, from a few hours to $\sim$ one day (see Extended Data Table 2 for exact values, and EDF 8 for light curve comparison). This could indicate that \target may be a longer time-scale ($\sim 25$ times longer than eRO-QPE1) QPE source. However, as discussed in detail in ``Quasi-periodic erupters and Swift J0230+28'' in the Supplementary Information, \target possesses some properties that differentiate it from, and makes it unique compared to, the four known sources of QPEs: i) the four QPE sources show a cool $\rightarrow$ warm$\rightarrow$ cool temperature evolution in each of the eruptions, while \target does not present any cooling at the decays of its eruptions, instead the temperature continuously increase until the source fades below the level of detectability; ii) the shapes of the eruptions in these four QPE sources are either nearly symmetric, or slightly asymmetric with longer decays than rises, while the rises in \target are $\sim30\%$ longer than the decays. 

A comparison between the general properties of \target, RNTs and QPEs is shown in Table \ref{tab:qpe_rnt}. In summary,  \target shows more similarities with QPEs sources than with RNTs. It shares many, but not all, of the properties of known QPE sources, and operates on an order-of-magnitude longer time-scale. 
Observationally, it resembles either a long time-scale QPE source, with slightly distinct properties, or the first example of a completely new class of transient with other members yet to be discovered. If the former case is assumed, one can investigate how \target's properties relate to those of the four known QPE sources. Fig.~5 shows that the recurrence time and the duration of the X-ray eruptions seem to be correlated, with a duty-cycle (duration/recurrence time) equal to $0.24 \pm 0.13$. A positive correlation also appears to be present between the recurrence time and amplitude of the eruptions, although only lower-limits on \target's amplitudes are known. However, the timing properties (duration and recurrence time) do not seem to be correlated with \mbh, perhaps indicating that the timescales of quasi-periodic, soft X-ray eruptions do not depend on \mbh.

Several models have been proposed to explain the repeating/recurrent phases of nuclear activity in RNTs and QPE sources, and those models can be roughly divided into two classes: those involving accretion-disc instabilities and those with smaller-mass bodies orbiting a MBH (i.e., extreme mass ratio inspirals; EMRIs). A more detailed discussion of each of these classes of models, and their strengths and weaknesses in explaining \target's properties, are presented in ``Physical models for \target'' in Supplementary Information, but in the following we summarize some of their aspects.

As for the accretion-disc driven models, some studies have proposed that instabilities associated with the inner accretion flow, from precession \cite{BarPett75,nixon12, Musoke2023,Liska2022} to distinct types of pressure/ionization instabilities \cite{MeyMeyH81,2020A&A...641A.167S,2022arXiv220410067S,2022ApJ...928L..18P,Kaur2023}, could produce quasi-periodic phases of high and low activity. However, the strong constraint on \target's host emission before the beginning of the eruptions as well as between eruptions ($\lambda_{Edd} \leq 0.002$) makes it extremely unlikely that a pre-existent standard thin disk is present, and the lack of any UV/optical variability makes accretion-disk instabilities an unfavorable interpretation for \target's eruptions, given that these models require either one or both of these to be observed (see ``Accretion disc instabilities'' in Supplementary Information for details).

The repeated partial tidal disruption (pTDE) of a star on a bound orbit about a MBH was proposed to explain both RNTs and QPEs. The repeated partial disruption of a main sequence star can explain the properties of RNTs \cite{Payne2021,Cufari2022,Wevers2023,Liu2023,Liu2023b}. In the case of \target, if the X-ray eruptions arise from accretion and the accretion efficiency is of the order $10\%$, the mass accreted per eruption is $\sim 10^{-4} - 10^{-5}M_{\odot}$ (see ``Eruptions Energetics'', in Methods). This implies that the mass lost by the star per orbit is a very small fraction of the total stellar mass, which would suggest that the pericenter distance of the star is extremely fine-tuned to coincide with the partial tidal disruption radius \cite{Guillochon2013}. This raises the question of how the star achieved such a fine-tuned distance, given our constraints on the beginning of the eruptions (see ``Constraints on the start of the eruptions'' in Methods). Furthermore, the shape of \target's eruptions is the opposite of the fast rise and longer decay expected from fallback accretion. A repeating pTDE of a white dwarf (WD), as proposed to explain the hour-long time-scale eruptions in QPE sources \cite{Zalamea2010,King2020}, can likely be excluded for \target given that a standard WD mass and corresponding radius \cite{Nauenberg1972} yields a tidal disruption radius that is close to a factor of 10 smaller than the direct capture radius of a $10^{6.6}$ \msol non-spinning black hole (see ``Repeating partial tidal disruption event'' in Supplementary Information for details).

After the discovery of the first QPE source, a series of alternative models related to distinct types of EMRIs have been proposed to explain the few-hours to a day, X-ray-only eruptions, which could, in principle, be extended to explain \target. These include collisions between a compact object or star and a pre-existing accretion disk/ADAF \cite{2021ApJ...917...43S,Linial2023,2022arXiv221008023L} (see ``Accretion disc – perturber interaction'' in Supplementary Information), the mass transfer from a single orbiting star undergoing Roche-lobe overflow around the MBH \cite{2022ApJ...941...24K,2023ApJ...945...86L} (see ``Stellar mass-transfer'' in Supplementary Information), or a pair of interaction stellar EMRIs \cite{2022ApJ...926..101M}(see ``Interacting stellar EMRIs'' in Supplementary Information), or alternatively the compression of stream clumps from a past TDE \cite{Guillochon2015,Coughlin2020c}(see ``Compressed reformed clumps from a past TDE'' in methods).
Several of these models can reproduce \target's features, while simultaneously suffering from inherent modeling uncertainties, degeneracies, or finely tuned parameters, as we discuss in detail in ``Extreme mass ratio inspirals (EMRIs)'' in Supplementary Information. Accurately constraining the characteristics of the physical system operating in \target thus remains a challenge.

Although the physical origin of QPEs, RNTs, and now \target are still the subject of debate, their discovery inaugurates a novel and exciting perspective on the study of transient events associated with MBHs, and \target has demonstrated the existence of a new timescale associated with these phenomena. In particular, if \target is a member of the same family as the four known QPE sources -- despite their slightly distinct properties -- and those originate from the same type of physical system, then the physical system's period needs to be scalable from a few hours to several days, i.e., more than two orders of magnitude, which would pose a strong constraint on potential theoretical models. 
Furthermore, there is a growing body of literature pointing toward these quasi-periodic X-ray eruptions being electromagnetic counterparts of EMRIs, with significant implications for the future of multi-messenger astrophysics.

The serendipitous discovery of \target also highlights exciting astrophysics that we are currently missing due to the lack of wide-field and time-domain X-ray surveys. The \textit{eROSITA} instrument \cite{Predehl2021} on the Spectrum-Roentgen-Gamma (SRG, \cite{Sunyaev2021}) space observatory has made significant progress in the field; however, its multiple visits of a field separated by four hours that are only revisited every six months makes it extremely unlikely to discover transients that vary on day timescales, 
such as \target. In the near future, the combination of the wide field of view and high cadence of \textit{Einstein Probe} \cite{Yuan2022} should in principle be able to discover more events similar to \target, although its shallow sensitivity combined with the likely very low rate of such event may result in no such discovery. In the future, only a deep and wide-field time-domain X-ray mission will be able to systematically discover a population of \target-like objects.

\clearpage

\begin{figure*}[htp!]
\centering
\includegraphics[width=0.85\textwidth]{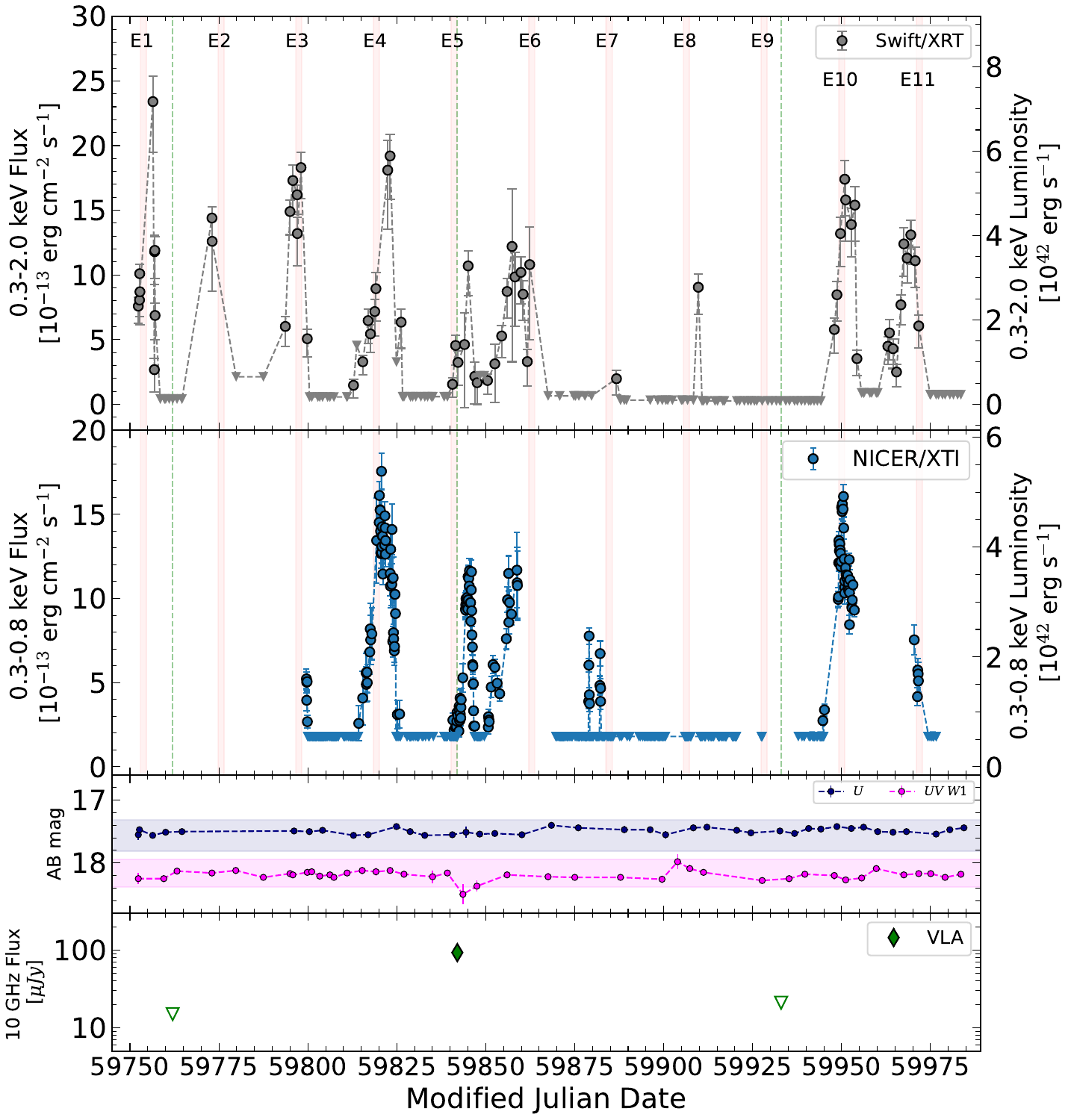}

\caption{\textbf{Light curves of \target.} {\bf Top}: \swift/XRT 0.3-2.0 keV flux and luminosity evolution. Stacked 3$\sigma$ upper limits between the eruptions are $2 \times 10^{-14}$ \fluxunits. {\bf Middle}: \nicer 0.3-0.8 keV flux and luminosity evolution. In both X-rays panels, circles are detections, reverse triangles are 3$\sigma$ upper limits of non-detections, and shaded pink regions indicate the $21.8^{+1.2}_{-0.5}$ days peak period found in the LSP analysis (see ``Time-resolved X-ray analyses'' in Methods). {\bf Bottom}: UV/optical and Radio light curves. \swift/UVOT UV $W1$ and $U$ bands are respectively dark blue and magenta points. The shaded region represents the $\pm 2 \sigma$ dispersion of the magnitude before the start of the X-ray eruptions (Dec 2021 to Jan 2022). Radio VLA observations are shown in green diamond (detection) and inverse triangles (non-detection upper limits), green dashed lines marks the epochs of the radio observations for reference.
 Error bars represent 1$\sigma$ uncertainties in all panels.} \label{fig:Fig1}

\end{figure*}

\newpage

\begin{figure*}[t]
    \centering
    \includegraphics[width=0.50\textwidth]{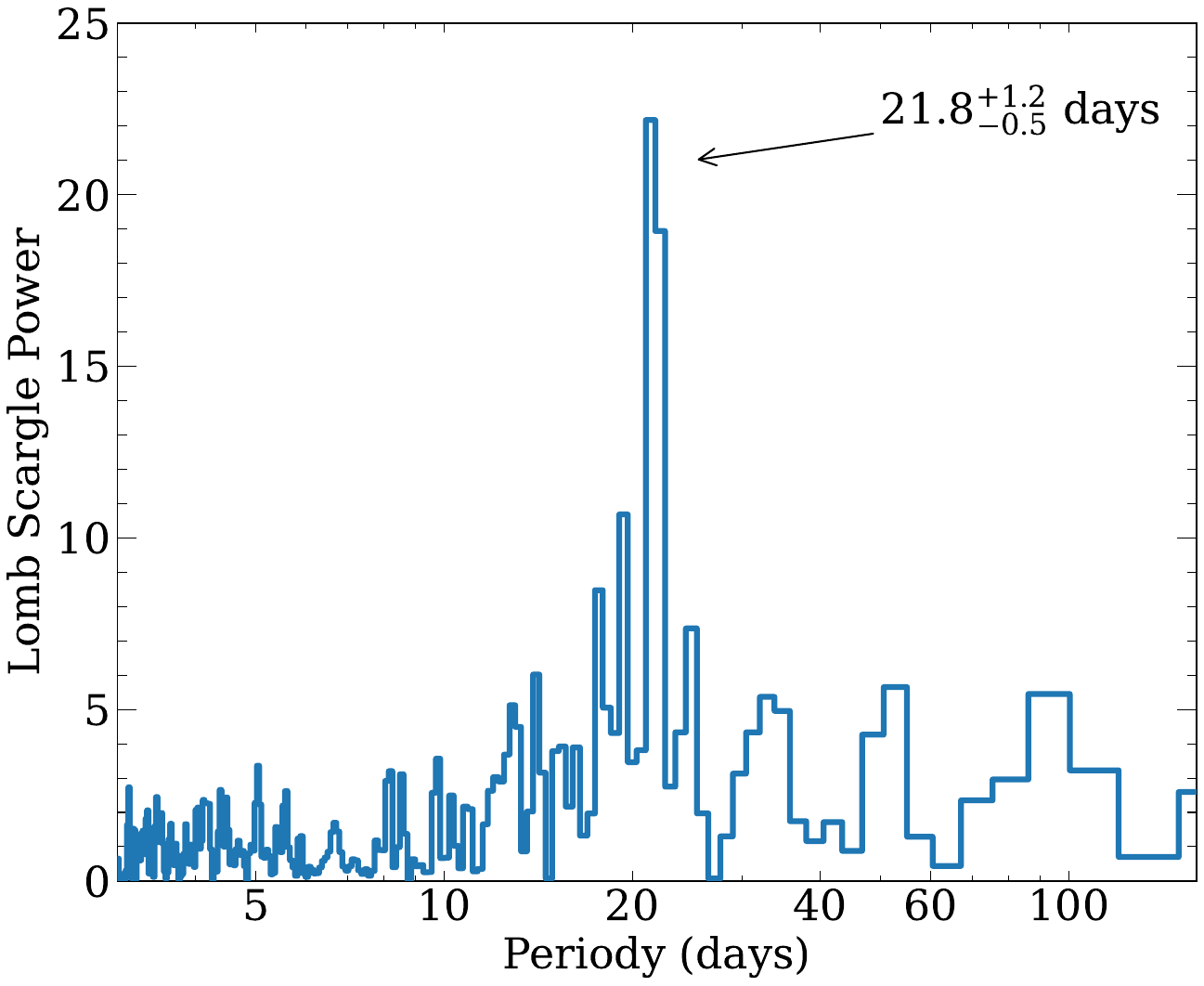}
  
    \caption{\textbf{Lomb-Scargle Periodogram (LSP) of \target light curve.} The two consecutive peak bins represent a $21.8^{+1.2}_{-0.5}$ days period.}
    \label{fig:LSP}
\end{figure*}

\begin{figure*}[b]
    \centering
    \includegraphics[width=\textwidth]{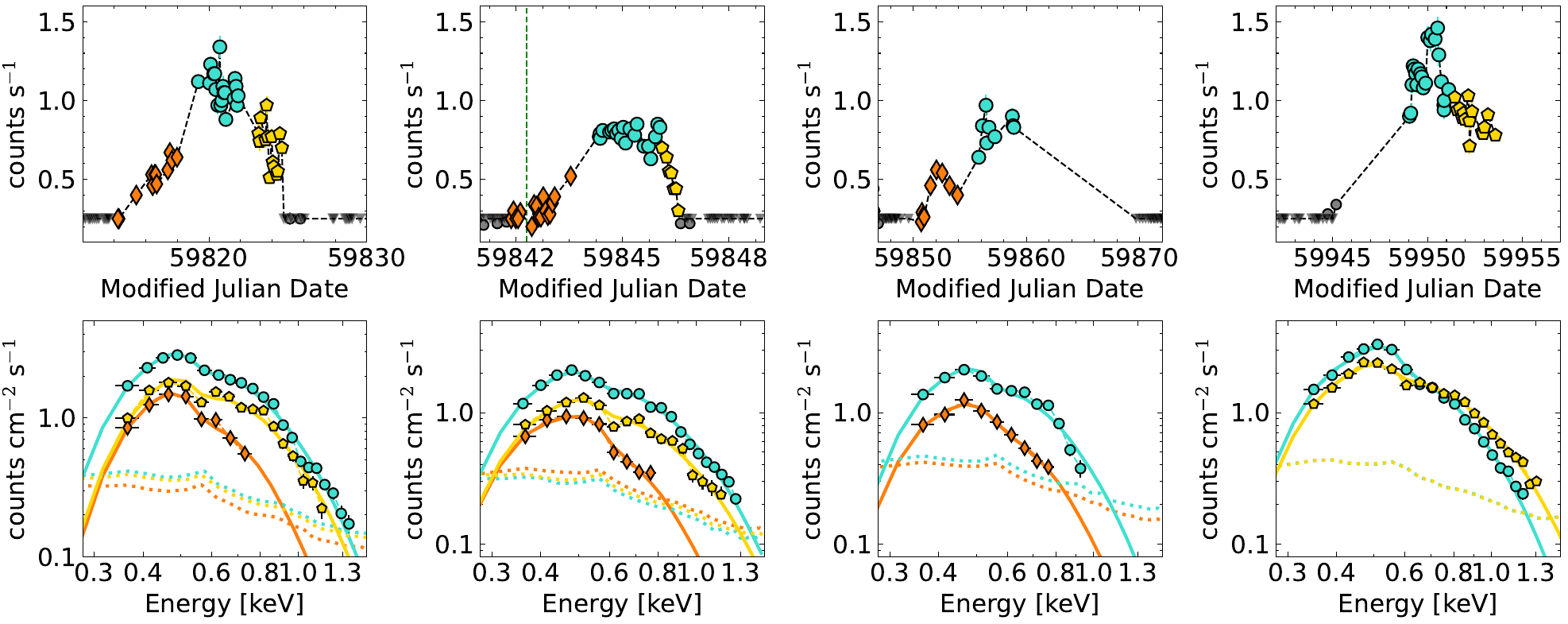}
  
    \caption{\textbf{\nicer Time-resolved spectroscopy.} {\bf Top}: Well-sampled \nicer light curves of four eruptions. Orange, cyan, and gold colors mark the observation stacked to create rise, peak, and decay spectra. The green dashed vertical line marks the epoch of the radio detection. {\bf Middle:} Observed spectra (markers) and best-fitting thermal model (continuous lines) and background spectra (dotted lines). Error bars represent 1$\sigma$ uncertainties in all panels.}
    \label{fig:Fig3}
\end{figure*}

\begin{figure*}
    \centering
    \includegraphics[width=\textwidth]{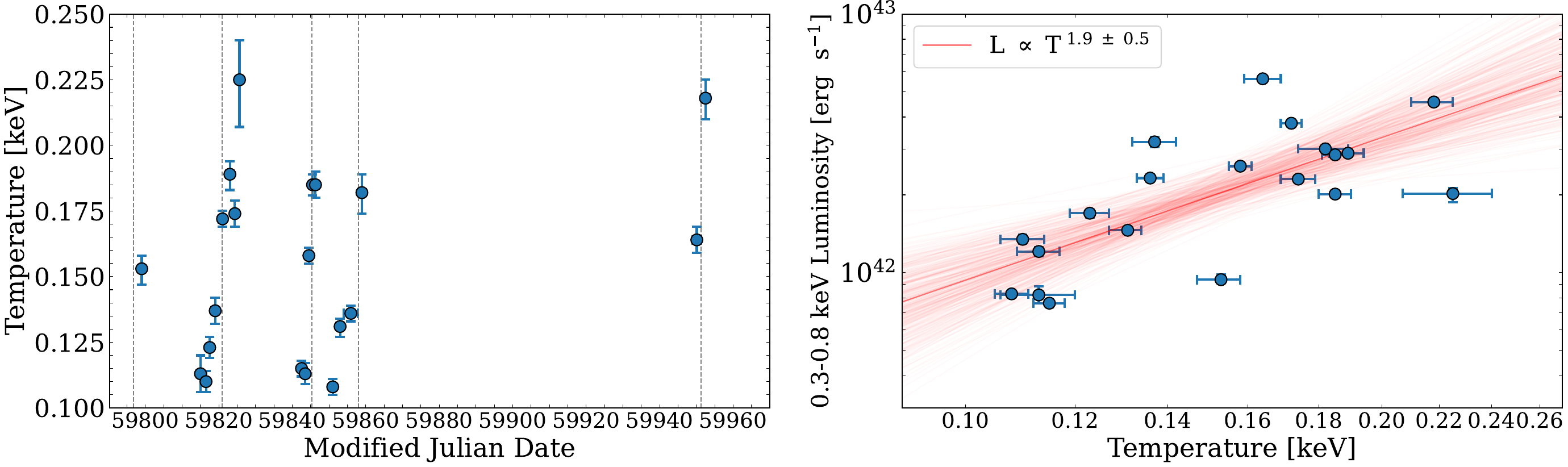}
    \caption{{\bf Time-resolved spectral properties}. {\bf Left}: The evolution of the temperature ($T_{in}$, using \texttt{diskbb}). {\bf Right}:  0.3-0.8 keV luminosity (\nicer) as a function of the temperature,  red lines are samples drawn from the posterior distributions of the best fit regression, $L \propto T^{1.9 \pm 0.5}$. Error-bars represent 1$\sigma$ uncertainty.}
    \label{fig:T_L}
\end{figure*}

\begin{figure*}[h!]
    \centering
    \includegraphics[width=0.75\textwidth]{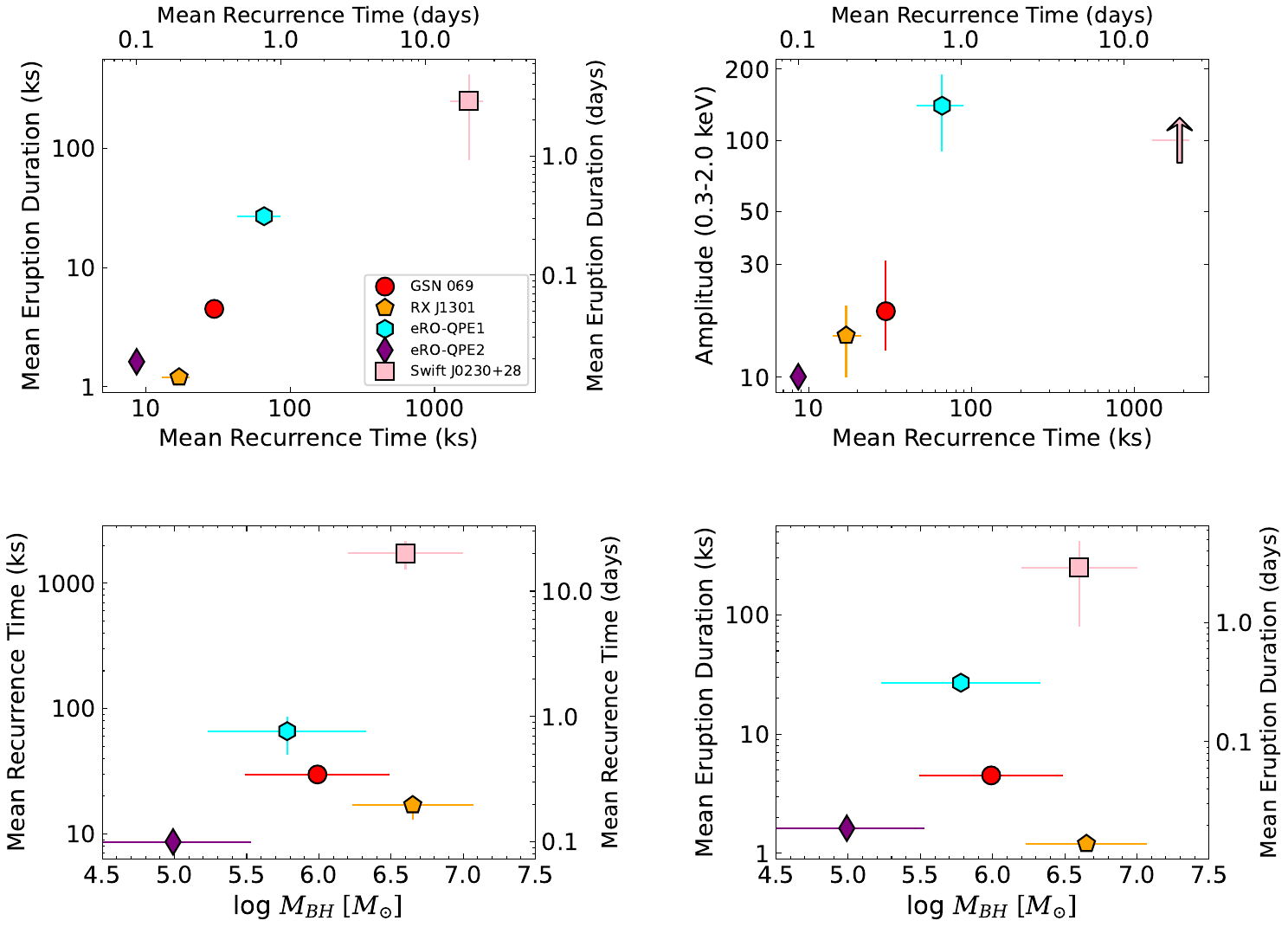}
  
    \caption{\textbf{Phase space diagrams for QPEs and \target.} \textbf{Upper Left:} mean QPE duration vs. mean recurrence time. \textbf{Upper Right:} Amplitude (0.3-2.0 keV band) vs. mean recurrence time. \textbf{Bottom Left:} Mean recurrence time vs. black hole mass (\mbh) derived from host-galaxy stellar velocity dispersion. \textbf{Bottom Right:} Mean Eruption Duration vs. black hole mass (\mbh). The top panels show some tentative correlations, such correlations are extended by at least an order of magnitude if \target is considered a QPE source. There is no correlation between the timing properties and the \mbh. This figure is based on \cite{Chakraborty2021}. The values are shown in Extended Data Table 2, uncertainties in the timing properties and amplitudes represent the full range of observed values, and the uncertainties in \mbh are 1$\sigma$.}
    \label{fig:QPE_relations}
\end{figure*}

\clearpage

\newpage

\begin{table}[ht!]
    \footnotesize
    \ttabbox[]{
    \begin{tabular}{|c|c|c|c|}
    \hline
    Properties                     & \target & \begin{tabular}[c]{@{}c@{}} QPEs \end{tabular} & RNTs \\ \hline
     \begin{tabular}[c]{@{}c@{}} Recurrent soft  \\ X-ray  eruptions \end{tabular} &  \begin{tabular}[c]{@{}c@{}} Yes   \\  \end{tabular}     & \begin{tabular}[c]{@{}c@{}} Yes   \\  \end{tabular}                                          & No\footnote{The outbursts in eRA J0456-20 evolve much smoothly than the eruptions in QPEs and \target. The periodic behavior of ASASSN-14ko is mostly found in the UV/optical bands, not so much in the X-rays. The X-ray spectra in all three RNT, are not soft, instead show hard X-ray emission from a corona. }                    \\ \hline
    Mean recurrence time           & $\sim$22 days & 2.4--18.5 hours                                         & 114--1200 days        \\ \hline
    X-ray spectra  & soft/thermal & soft/thermal                                      &  \begin{tabular}[c]{@{}c@{}}soft/thermal + corona \\ or corona only \end{tabular}      \\ \hline
     \begin{tabular}[c]{@{}c@{}} X-ray luminosity at peak\footnote{In the 0.3-2 band for \target and QPEs, and in the 0.3-10 keV band for RNTs.} \\ at peak (\ergs) \end{tabular}   &  $\rm{few} \times 10^{42}$          &        $10^{42}-10^{43}$             &    $> \rm{few} \times 10^{43} $   \\ \hline
    UV/optical emission            & \text{\sffamily x}     & \text{\sffamily x}  & \checkmark            \\ \hline
     \begin{tabular}[c]{@{}c@{}} UV/optical luminosity at peak \\  (\ergs) \end{tabular}           &  $< 3 \times 10^{42}$\footnote{Based on the lack of variability in \swift/UVOT bands, see ``UV/optical and radio counterparts'', in methods.}  &$ \ll 10^{43 } $\footnote{Based on the lack of variability in \swift/UVOT and/or \xmm/OM instruments from the host level.}          &    $\geq 10^{44}$                                                                       \\ \hline
      Light curve shape         & \begin{tabular}[c]{@{}c@{}c@{}} slight asymmetric  \\ (rises $\sim$30$\%$ longer \\ than decays)\end{tabular}       &    \begin{tabular}[c]{@{}c@{}c@{}} symmetric or  \\  slight asymmetric \\ (longer decays than rises) \end{tabular}     &       \begin{tabular}[c]{@{}c@{}c@{}} complex and \\  varied \footnote{ASASNN-14ko and AT2018fyk have TDE-like (rapid rise, decay $\rm{several} \ \times$ longer). eRA J0456-20 show much longer rises than decay.   } \end{tabular}   \\ \hline
        \mbh   & $10^{6.6}$ \msun     & $10^{5}$--$10^{6.6}$ \msun                                              & $10^{7.1}$--$10^{7.7}$ \msun                                                                                       \\ \hline
     \hline

    \end{tabular}
      
    \caption{ \footnotesize \textbf{ General properties of \target as compared to quasi-periodic eruption sources (QPEs) and repeating nuclear transients (RNTs).} QPEs are GSN 069 \cite{Miniutti2019}, RX J1301 \cite{Giustini2020}, eRO-QPE1 and eRO-QPE2 \cite{Arcodia2021}. RNTs are ASASSN-14ko \cite{Payne2021},
    eRA J0456-20 \cite{Liu2023} 
    and AT2018fyk \cite{Wevers2023}.}
    \label{tab:qpe_rnt}
    }

\end{table}
\section*{{\Huge Methods.}}

\section{\Large{\bf Observations and Data Analysis}}\label{supsec:data}
This work is based on novel data acquired by 5 different telescopes/instruments, as well as in archived data, across the entire electromagnetic spectrum (namely Radio, Infrared, optical, UV and X-rays). Below, we describe the data and their relevant reduction and analysis procedures. Throughout this paper, we
adopt a standard $\Lambda$CDM cosmology with H$_{0}$ = 67.4 km~s$^{-1}$~Mpc$^{-1}$, $\Omega_{m}$ = 0.315 and $\Omega_{\Lambda}$ = 1 - $\Omega_{m}$ = 0.685 \cite{planck}. Using the Cosmology Calculator \cite{Wright2006} \target's redshift (z) of 0.036 corresponds to a luminosity distance of 165 Mpcs.

\subsection{\swift/X-Ray Telescope (XRT)}\label{data:xrt}
The field containing the position of \target was observed by \swift between December 2021 and January 2022 following the discovery of the supernova SN2021afkk located 4.25 arcmin from the position of \target. During that time X-rays were not detected from the position of \target with a 0.3-2 keV flux upper limit of $2\times10^{-14}$ \fluxunits. After a gap of 164 days \swift started again monitoring the field of view on 22 June 2022, making the first detection of the new X-ray source at the position RA, Dec: 37.57140$^o$, 28.60124$^o$, with an uncertainty of 3.4\arcsec (radius, 90\% confidence). While the monitoring continues at the time of writing of this manuscript we include all data taken until 8 February 2023 (MJD 59983).

We started XRT data analysis by downloading the data from the HEASARC public archive. We extracted the cleaned event files by running the {\tt xrtpipeline} on each observation ID (obsID). 
For each obsID we run a detection pipeline \cite{Evans2014}, using a circular extraction region centered on (02:30:17.1,+28:36:04.5) (J2000.0 epoch) with a radius of 47\arcsec and a background using an annulus centered on the source position with an inner and an outer radii of 80\arcsec and 250\arcsec, respectively. To consider a detection, we require that the source was detected above the background at a confidence of at least $3\sigma$, in a Bayesian framework \cite{Kraft1991}, below that we consider as a non-detection.

\subsubsection{XRT count rate to flux and luminosity conversion}\label{sec:xrtfactor}
We followed the procedure below to covert from observed 0.3-2.0 keV background-subtracted count rate to observed 0.3-2.0 keV flux and luminosity.
\begin{enumerate}
    \item First, we extracted a combined source spectrum using obsIDs where the source was detected above $3\sigma$.
    \item Then we combined the corresponding individual exposure maps and used it to compute a combined ancillary response file by following the steps outlined on XRT's data analysis webpages here: \href{https://www.swift.ac.uk/analysis/xrt/exposuremaps.php}{https://www.swift.ac.uk/analysis/xrt/exposuremaps.php} and \href{https://www.swift.ac.uk/analysis/xrt/arfs.php}{https://www.swift.ac.uk/analysis/xrt/arfs.php}.
    \item Using {\tt ftool} {\tt ftgroupha} we grouped the spectrum using the optimal binning scheme of \cite{Kaastra2016} with the additional requirement to have at least 20 counts per spectral bin
    \item The resulting combined spectrum was then fit with a thermal model, i.e., {\tt tbabs*zashift*diskbb} in {\tt XSPEC}. {\tt tbabs}'s column was fixed at the MilkyWay value (7.2 $\times 10^{20}$ cm$^{-2}$) using HEASARC $N_H$ calculator: \href{https://heasarc.gsfc.nasa.gov/cgi-bin/Tools/w3nh/w3nh.pl}{https://heasarc.gsfc.nasa.gov/cgi-bin/Tools/w3nh/w3nh.pl}. This gave an acceptable $\chi^{2}$/degrees of freedom (d.o.f) of 12/11. The mean 0.3-2.0 keV background-subtracted count rate in this combined spectrum was 1.73$\times$10$^{-2}$. The observed flux and luminosity were 4.3$\times$10$^{-13}$ \fluxunits and 1.5$\times$10$^{42}$ \ergs, respectively. Based on this we derive a background-subtracted 0.3-2.0 keV count rate to flux (luminosity) scale factor of 2.5$\times$10$^{-11}$ \fluxunits/count s$^{-1}$ (9$\times$10$^{43}$ \ergs/counts s$^{-1}$).
\end{enumerate}

\subsubsection{Estimating the XRT X-ray upper limit}\label{sec:XRT_upperlim}
The source is not detected in several observations taken during the low-flux state between the eruptions. We estimate the flux upper limit as follows:
\begin{enumerate}
    \item For given $N$ consecutive non-detections (source less than $3\sigma$ above background) we extract a combined 0.3-2.0 keV  image for the $N$ OBS ID's.
    \item Using {\tt ximage}'s {\tt sosta} functionality we estimated the count rate upper limit for each of groups of $N$ non-detections and obtained the $3\sigma$ flux upper-limit using the scaling factor determined in section \S\ref{sec:xrtfactor}. We attribute the combined upper-limit of the grouped observations, for each of the  OBS ID.
    \item We estimated the 0.3-2.0 keV flux upper limit for the combined image of all non-detections, which of 7$\times$10$^{-4}$ counts/s which translated to roughly 2$\times$10$^{-14}$ \fluxunits.
    \item We performed a similar procedure in the 2-10 keV band to estimate an upper limit for a AGN/corona emission. We obtained 5$\times$10$^{-4}$ counts/s which translate (assuming a power-law spectrum with $\Gamma$ = 2) to a 2-10 keV flux upper limit of 3$\times$10$^{-14}$ \fluxunits.
\end{enumerate}

The resulting XRT light curve showing flux (luminosity) versus time is shown in upper panel of Fig.~1, detections are shown in solid circles with error-bars representing 1$\sigma$ uncertainty, while non-detections are shown as inverse triangles.

\subsection{\swift/UVOT}\label{supsec:uvot}

All the data were processed with \texttt{heasoft} v6.29c. We use the \texttt{uvotsource} package to extract the UVOT photometry measurements, using an aperture of 5\arcsec. We include observations prior and during the X-ray transient. Given that most observations were performed in 
the  `filter of the day' configuration not all the UVOT filters are available for all epochs. Photometry was corrected for Galactic extinction E(B-V) = 0.086 
\cite{Schlafly2011}. There is no statistically significant variability in any of the UVOT filters. We show the well sampled UV W1 and U filter light curves in the bottom panel of 
Fig.~1.

\subsection{\nicer}\label{supsec:nicer}
Following \swift/XRT's detection of highly variable X-ray emission from \target \nicer started a monitoring program as part of Director's Discretionary Time (DDT). \nicer observations started on 23 June 2022 (MJD 59753) and continue at the time of writing of this paper. Here we include data taken until 1 February 2023 (MJD 59976). \nicer's observing cadence varied during this time with 1-2 exposures per day in the epochs between the eruptions and several exposures per day during the eruptions. The individual exposures varied in length between 200 and 1000 seconds. Stacked image from the XRT ($\sim$ 160 ks, see top left panel EDF 5) show that only \target is
detected within the NICER field of view, enabling detailed analyses of NICER spectra with
no concerns for contamination by other sources.

We started our data analysis by downloading the data from public HEASARC archive (\href{https://heasarc.gsfc.nasa.gov/docs/archive.html}{https://heasarc.gsfc.nasa.gov/docs/archive.html}). 
We essentially followed the same reduction procedures as outlined in-detail in \cite{Pasham2022cmc,2018cowpasham}. 
The cleaned events lists were extracted using the standard NICER Data Analysis Software
(HEASoft 6.29) tasks \texttt{nicercal}, \texttt{nimpumerge}, and \texttt{nicerclean} The latest NICER calibration release xti20221001 (01 October 2022) was used.
The cleaned event
files were barycenter-corrected using the barycorr ftools task. \target’s coordinates (J2000.0): (02:30:17.1, +28:36:04.) were used along with \textit{refframe=ICRS} and
\textit{ephem=JPLEPH.430}. The Good Time Intervals (GTIs) were extracted with the \texttt{nimaketime}
tool using the default filters: \textit{nicersaafilt=YES}, \textit{saafilt=NO}, \textit{trackfilt=YES}, \textit{ang dist=0.015},
\textit{st valid=YES}, \textit{cor range=``*-*”}, \textit{min fpm=38}, \textit{underonly} \textit{range=0-200},
\textit{overonly} \textit{range=“0.0-1.0”}, \textit{overonly expr=“1.52*COR SAX**(-0.633)”}, Conservative \textit{elv=30}
and \textit{br earth=40} were used to avoid optical loading by reflected light.

\subsubsection{Converting from \nicer count rate to luminosity}\label{subsec:nicercrtolum}
We computed the fluxes on a per GTI basis using the following procedure:
\begin{enumerate}
    \item First, we extracted time-resolved \nicer spectra of the source and estimated background spectra using the 3c50 model \cite{Remillard2022}. Corresponding response files (arf and rmfs) were extracted using the {\tt nicerarf} and {\tt nicerrmf} tools.
    \item Because the 0.8-1.3 keV bandpass is occasionally dominated by the systematic residuals we fit each of the above spectra in the 0.3-0.8 keV with a thermal model ({\tt tbabs*} {\tt zashift*}{\tt diskbb}) in {\tt XSPEC}\cite{xspec}. Following the recommendation by \cite{Remillard2022} we only consider spectra where the background-subtracted source count rate was greater than 0.25 counts/s \and had a signal-to-noise ratio (source background-subtracted over background) higher than three.  
    \item Using the above spectral fitting we estimated an epoch-dependent background subtracted 0.3-0.8 keV count rate--to--observed luminosity conversion factor which was used to compute the observed luminosities for each GTI. All GTIs with a background subtracted 0.3-0.8 keV count rate of less than 0.25 count/sec or signal-to-noise ratio was less than 3 were assigned an 0.3-0.8 keV flux upper limit of $1.79 \times 10^{-13}$ \fluxunits  (triangles in the panel middle panel of Fig.~1).

\end{enumerate}

For all \nicer spectral fitting in this work, we have used  \textit{optimal} spectral binning scheme \cite{Kaastra2016}, $\chi^2$ statistics, as well as an additional 1.5\% systematic uncertainty. Similarly to the \swift/XRT fitting we always assume Galactic absorption value following HEASARC's $N_H$ calculator.

\subsection{Very Large Array (VLA)}\label{supsec:radio}

We observed \target\ with NSF's Karl G. Jansky Very Large Array (VLA) on 2022 July 2 (MJD 59762), 2022 September 20 (MJD 59842), and 2022  December 20 (MJD MJD 59933). The VLA was in its highest-resolution A configuration during the first observation, its lowest-resolution D configuration during the second, and the intermediate C configuration during the third. All observations had one hour of exposure time and were conducted at a mean frequency of 10 GHz. The data were reduced in CASA \cite{CASA}
using standard procedures. We additionally performed amplitude and phase self-calibration on the September data. There was no detection at the position of the source in the first and third observations, with 3$\sigma$ upper limits, of 15 $\mu$Jy and 25 $\mu$Jy, respectively. However, in our second observation on 2022 September 20, we detect an unresolved radio source, with a flux density of $93\pm7$ $\mu$Jy  (13 $\sigma$ detection).

\subsection{Optical Spectra}\label{supsec:op_spec}
\subsubsection{Keck-II/ESI}

On 2022 October 24.3, we obtained a medium resolution spectrum of the host galaxy nucleus using the Echellette Spectrograph and Imager (ESI; \cite{Sheinis2002}) on the Keck II telescope. We used the Echelle mode and a slit width of 0.5\arcsec, which gives an instrumental broadening of $\sigma_{\rm inst} = 15.8$\,$\rm km\,s^{-1}$. The exposure time was 15\,min. The median signal-to-noise ratio (S/N) from 4500\,\AA\ to 5600\,\AA\ is 9.

\subsubsection{VLT/X-shooter}
The host galaxy was also observed under DDT programme 110.2599 on 2022 November 17 with X-shooter \cite{Vernet2011} mounted on the Very Large Telescope (VLT) Unit Telescope 3 (Melipal). Slit widths of 1/0.9/0.9 arcsec were used for the UVB/VIS/NIR arms, providing a spectral resolution R = 5400/8900/5600, respectively, covering the spectral range 3500--24800 \AA. Exposure times were 2$\times$1497/1411/1200 seconds in the UVB/VIS/NIR arms, with an on-slit nod to facilitate sky line subtraction in the NIR arm. We reduce the data using the standard pipeline recipes within {\tt esoreflex}. For the UVB and VIS arms, we use recipes for stare observations to increase the S/N. Strong, extended nebular emission from the host galaxy complicates the sky subtraction, leaving strong residuals and oversubtractions, especially in the Balmer lines. To measure the line profiles and fluxes for the diagnostic diagrams, we therefore use extractions without subtracting the skylines. In order to measure the velocity dispersion, where the emission lines are masked, we use sky-subtracted extractions. The continuum normalized spectrum in the H$\beta$+[O\,\textsc{iii}] and H$\alpha$+[N\,\textsc{ii}] bands are shown in the bottom panel of EDF 5.

\subsubsection{Continuum and emission line fitting}

We measured the velocity dispersion $\sigma_\ast$ with the penalized pixel-fitting (\texttt{pPXF}) software \cite{Cappellari2004, Cappellari2017}, which fits the absorption line spectrum by convolving a template stellar spectral library with Gauss-Hermite functions.
We used the ELODIE v3.1 high resolution ($R=42000$) library \cite{Prugniel2001, Prugniel2007}, and masked wavelength ranges of common galaxy emission lines, hydrogen Balmer lines, telluric regions, and an instrument artifact feature at an observer-frame wavelength of $\sim 4510$\,\AA. 
Following previous works \cite{Wevers2017, Wevers2019_Mbh, French2020, Yao2022}, we performed 1000 Monte Carlo (MC) simulations to robustly determine $\sigma_\ast$. 

The Keck spectrum has a relatively low signal-to-noise ratio (a factor of 2 smaller than that of the X-shooter data), as well as covering a limited wavelength range (4500 -- 5600 \AA\xspace compared to 3500--9000 \AA\xspace for X-shooter). Both of these factors can introduce significant systematic uncertainties related to template mismatching and spurious results in the MCMC step. For these reasons, we use the X-shooter data to measure the velocity dispersion and estimate the black hole mass.

We measure a velocity dispersion of $\sigma_{\ast} = $ 87 $\pm$ 2 km s$^{-1}$ which along with the \mbh-$\sigma$ relation \cite{Gultekin2009} allowed us to estimate a black hole mass of log \mbh = $6.6 \pm 0.4$ \msol.

Emission/absorption line fluxes and equivalent widths (EW) were measured using continuum-normalized versions of the ESI and X-shooter spectra. We include a low order polynomial for the continuum, and single Gaussian components for each emission line, including H$\beta$, H$\alpha$, [O\,\textsc{iii}]$\lambda 5007$, [O\,\textsc{i}]$\lambda 6300$, and the [N\,\textsc{ii}] and [S\,\textsc{ii}] doublets. No broad emission line components are evident. Typical line widths for the narrow components are 150--200 km s$^{-1}$, and some lines show weak asymmetries (where the lines are skewed to the blue wing). The line measurements and their ratios are used to locate the host galaxy nucleus on diagnostic BPT and WHAN diagrams, as shown in the top and middle panels of EDF 6. From the X-shooter spectrum we also measure the H$\delta$ Lick absorption index \cite{Worthey1994}, which can be used to identify quiescent Balmer strong and E+A (post-merger) galaxies. We plot \target among the SDSS population (grey background points), and also include measurements from the ZTF TDE sample \cite{Hammerstein_21} and QPE host galaxies \cite{Wevers2022} in the bottom right panel of EDF~6. \target is classified as a QBS galaxy, which comprises less than 2.3\% of the total galaxy population. This is consistent with the over-representation observed in the QPE host population \cite{Wevers2022}.

\section{\Large{\bf Time-resolved X-ray analyses}}\label{supsec:spec_analyses}

\subsection{X-ray light curve}
\label{subsec:timming}

\target showed phases of high (detection) and low (non-detections) activity over the course of our \swift/XRT monitoring (Fig.~1). A quick visual inspection suggests that these eruptions repeat roughly in the order of $\sim 3$ weeks. To test for a quasi-periodicity we extracted a Lomb-Scargle Periodogram (LSP) of the \swift/XRT 0.3-2.0 keV light curve following the description in  \cite{scargle,lspnorm}, which is shown in Fig.~2. Not surprisingly, the LSP shows two consecutive bin peaks around 21-23 days, with the bin with highest power at 21.8 days. The combined two bins their FWHM result in a peak period of $21.8^{1.2}_{-0.5}$ days. Therefore, the LSP confirms the quasi-periodic nature of \target's eruptions.

The eruptions in \target are apparently asymmetric, with slower rises than decays. In order to quantify such asymmetries, we fit the six best-sampled eruptions, namely around \textit{E3, E4, E5, E6, E10, E11}, with an asymmetric Gaussian function $G(\mu, \sigma_+, \sigma_-)$) \cite{Barlow2004}, where $\sigma_{\pm}$ are the equivalent positive/negative $\sigma$. We use either the \nicer or the \swift light curve depending on which had a better sampling of the eruption's shape at the given epoch.
In the left panel of EDF 5 we show the best-fitted asymmetric Gaussian to the six eruptions. In the right panel, we quantify the asymmetry in the eruptions by plotting the $\sigma_+/\sigma_-$ ratios: which vary between $0.60 \pm 0.10$ to $0.90\pm0.05$, and show a median value of $\left \langle \sigma_+/\sigma_-   \right \rangle \approx 0.7$, confirming that the eruptions are asymmetric, with slight slower rises than decays. The best-fitted parameters for the profile fitting are shown in Supplementary Information Table 1.

\subsection{X-ray spectra}
\label{subsec:T}
The high count rate resulting from the high effective area of \nicer, allows for detailed time-resolved spectral analyses, not possible with \swift/XRT. First, we stack \nicer data of detected epochs in bins of $\sim$ 1 day, each spectra/bin having between 300 and 9000 counts. The energy range in which the source is detected above the background varies depending on the phase of the eruption; however, at all epochs, it is detected at least up to 0.8 keV; hence to measure luminosities and temperatures, we perform the fitting procedures described in \ref{subsec:nicercrtolum} in the 0.3-0.8 keV band. The temperature varies by a factor of two, while the luminosity (in the 0.3-0.8 keV band) varies by a factor of 10 between $5 \times 10^{41}$ \ergs and $5 \times 10^{42}$ \ergs.

The resulting values are shown in Extended Data Table 2. The evolution of the temperature as a function of time is shown in the left panel of Fig.~4; the vertical dashed line shows the peak of each eruption (as fitted in \ref{subsec:timming}), in the left panel, shows luminosity as a function of temperature. At each eruption, the temperature starts $\sim$ 100 eV and continuously increases up to $\sim$ 200 eV at the decays; interestingly, the temperature does not peak at the luminosity peak but instead peaks at the decay phases.
Despite the absence of a clear ``hotter when brighter" trend in the temperature evolution, we employed the \texttt{linmix} package \cite{Kelly2007} to fit the relationship between the two parameters and still found a positive correlation in the form of $L \propto T_{in}^{ 1.9 \pm 0.5}$, although the large scatter is likely driven by the hotter temperatures at the decay phases.

To increase the signal-to-noise ratio and probe the spectra shape at energies higher than 0.8 keV we divided each of the four eruptions probed by \nicer (namely, those around \textit{E4, E5, E6} and \textit{E10}) in three phases, rise, peak and decay, and produce one stacked spectra for each of the phases of the eruptions, the resulting spectra have between $\sim$ 2000 and $\sim$ 20000 background-free counts, an are detected above background at higher energies, up to 1.4 keV for some peak and decay spectra. We fit the resulting high signal-to-noise ratio spectra assuming the same model as before (i.e., \texttt{TBabs$\times$zashift$\times$diskbb}) which results in $\chi^2$/$d.o.f$ varying between $\sim 1.1$ and $\sim 2.5$, with residuals present both at the softest end of the spectra and around 1.0 keV for some peak and rise spectra. We then added an intrinsic absorption component component at the redshift of the source (i.e. \texttt{TBabs$\times$zTBabs$\times$zashift$\times$(diskbb})), such absorption takes care of the residuals at the softer energies, and results in better fitting for all spectra, with $\chi^2$/$d.o.f$ between $\sim 0.9$ and $\sim 1.9$, the best-fitted intrinsic column density was $N_H \approx (1-3) \times 10^{20}$ cm$^{-2}$ in all spectra.

We show the stacked spectra and best fit model in Fig.~3, while the respective residuals are shown in EDF 3. In a few of the spectra, e.g. some of the peak phases show an absorption-like residuals around 1.0 keV while in others, e.g. some of the decay spectra, the residuals are randomly distributed. A detailed study on whether such residuals are intrinsic to the source, e.g. an absorption line like the one detected in TDE ASASSN-14li \cite{Kara2018} or merely an instrumental/systematic residual from the \nicer instrument is beyond the scope of this study, and will be presented in a separated study in the case the former is confirmed. 

We also tested alternative models to the continuum, changing \texttt{dikbb} for a phenomenological power-law (\texttt{powerlaw}) or a single temperature blackbody (\texttt{blackbody}) results in a worse fit. 
A thermal bremsstrahlung model (\texttt{bremss}), however, results in just a slightly worse fit as compared to \texttt{diskbb}, in terms, $\chi^2$/$d.o.f$, in some of the spectra, and has similar fitting statistics in others. The ratio between the best-fitted bremsstrahlung's plasma temperature ($T_p$) and the inner disk temperature ($T_{in}$) is consistently $T_p/T_{in} \approx 2$, i.e. $T_p$ varied between $\sim$200 eV and $\sim$ 400 eV, while the same relation with the luminosity is observed. We also attempt to fit a model with two continuum components, by adding a powerlaw to the thermal emission (i.e., \texttt{TBabs$\times$zTBabs$\times$zashift$\times$}\texttt{(diskbb} \texttt{+powerlaw})), however, this does not improve the fit quality and the best-fit powerlaw normalization is in most cases negligible.
Given that the better overall fit is achieved with \texttt{TBabs$\times$zTBabs$\times$} \texttt{zashift$\times$diskbb} which is the final model for \target, and report in Extended Data Table~1 the best fitted parameters of the stacked spectra.

\subsection{Eruptions Energetics}

We estimated the energy released by individual eruptions by integrating their Swift/XRT light curves, an order-of-magnitude estimate resulting in $\sim$ few $\times 10^{48}$ erg per eruption depending on the duration/amplitude of the eruption. Assuming a 10\% efficiency ($\alpha$ = 0.1) in the mass to energy conversion and assuming that the eruptions are powered by accretion, results in few $\times 10^{-5}$ \msol accreted per eruption . If the eruptions started somewhere after 8 January 2022 (MJD 59587, see ``Constraints on the start of the eruptions'') then the maximum total  mass accreted is $< 10^{-3}$ \msol.

\section{UV/optical and radio counterparts}
\label{sec:sed}

Our three VLA observations, show two non-detections and one detection, the radio detection coincides with the rise of \textit{E5} eruption, while the non-detections coincide with the X-ray quiescent epochs, one between \textit{E1} and \textit{E2} and the other between \textit{E9} and \textit{E10} as can be seen by Fig.~1. The three radio images are shown in EDF 4.

The \swift/UVOT shows no variability that is more significant the $2\sigma$ from the pre-eruption host galaxy level, as can be seen in Fig.~1. From the observations we can compute an upper limit on the UV/optical variability, the UV~W1 in particularly allows for the deepest constraint. The derived observed upper limit of $\nu L_{\nu}(UV~W1)\leq 1.8 \times 10^{42}$ \ergs. In TDE studies the UV/optical integrated emission ($L_{BB}$) is estimated from the fit of a blackbody function to the UV/optical broad-band SED, and the UV/optical to X-ray luminosity ratio ($L_{BB}/L_{X}$) is the parameter of interest, to study the SED shape. Assuming a typical temperature found in UV/optical component of TDEs, around 20,000  Kelvin \cite{Hammerstein_21}, the UV~W1 upper limit translates into a $L_{BB} \leq 3\times10^{42}\,{\rm erg\,s^{-1}}$. This means, that at the peak of the X-ray eruptions $L_{BB}/L_{X} \leq 0.5$. In the stacked spectra fitting (see \S\ref{supsec:spec_analyses}) we found evidence for a very small intrinsic column density at the maximum $\sim 3\times10^{20}$  cm$^{-2}$, assuming a standard gas-to-dust ratio ($N_H = 5 \times 10^{21} \ \rm{cm}^{-2} \times E(B-V)$,\cite{Predehl1995}), which translates to a maximum dust extinction of $E(B-V) \approx 0.06$, assuming  a standard extinction law \cite{Calzetti_2000}, the intrinsic extinction-corrected emission in the $UV~W1$ band can be at maximum a factor of $\sim40\%$ higher. This just slightly increases the upper limits on the $UV~W1$ and integrated UV/optical emission to $\nu L_{\nu}(UV~W1)\leq 2.6 \times 10^{42}$ \ergs and $L_{BB} \leq 4.2\times10^{42}\,{\rm erg\,s^{-1}}$, respectively, meaning that extinction cannot be the cause of UV/optical faintness, such faintness is an intrinsic characteristic of \target eruptions.


\section{The host galaxy}\label{supsec:host}

The position derived from XRT for \target is 0.2\arcsec away from the photometric center of the nearby spiral galaxy 2MASX J02301709+2836050 located at 165 Mpcs. The galaxy shows prominent blue spiral arms and a redder bright nuclear core (see the top right panel in EDF 5). 

The host galaxy does not appear in large AGN catalogs \cite{veron_agn,Milliquas}. There is no previous X-ray detection at the position of this galaxy: the X-ray upper limit server \cite{up_lim_server} returns only upper limits, from \xmm Slew Survey, ROSAT all-sky survey (RASS) and previous observations by \swift/XRT (see exact values in section \S\ref{sec:constraint_begin}). The galaxy shows no excess in the infrared bands that could indicate the presence of a hot dust emitting `torus': the Wide-field Infrared Survey Explorer (WISE, \cite{Wise}) infrared  W1 - W2 color $\sim$ 0.1  does not pass standard AGN selection criteria \cite{Stern}, instead is consistent with pure stellar population emission. The NeoWISE \cite{neowise} light curves from 2014 to 2021 show no significant variability ($< 1\sigma$). Our recently obtained optical spectra show no signs of broad emission lines. Together, these multi-wavelength properties exclude the existence of a bright AGN (e.g., Seyfert-I or Quasar-like) in \target's host.

The narrow emission lines measured on the ESI and X-shooter spectra can be used to locate the host galaxy nucleus on diagnostic BPT \cite{Baldwin1981} and WHAN \cite{Cid2011} diagrams to investigate the ionization mechanism producing the lines, shown in the top and bottom panels of EDF 6. \target's host is located above the theoretical upper limit for pure star-forming galaxies \cite{Kewley2001} -- meaning an additional ionizing mechanism is necessary to produce these line ratios -- and around the empirical separation between Seyfert and low-ionization nuclear emission-line region (LINER) \cite{Cid2010,Kewley2006} on both [N\,\textsc{ii}] and [S\,\textsc{ii}] diagrams. The WHAN diagram \cite{Cid2011} can be used to classify the nucleus further: the low EW H$\alpha$ results in a weak-AGN (wAGN) classification. In general, the multi-wavelength properties agree that \target's may host (or have hosted) a low-luminosity AGN (LLAGN).
We gathered archival photometric data on the host galaxy: in the UV and optical from the UVOT (see \ref{supsec:uvot}), in the optical and near IR ($g$, $r$, $i$, $z$ and $y$ bands) from the PAN-STARRS survey \cite{Kaiser2002}, near IR ($K$, $H$, and $J$ bands) from the 2MASS survey \cite{Cutri2003} and in the mid-IR ($W1$, $W2$, $W3$, and $W4$ bands) from the WISE survey \cite{Cutri_13}. While gathering all the photometric data, we use the values extracted from aperture sizes as close to the Kron radius (12'') as possible for all the surveys. To estimate the host galaxy properties, we model the resulting spectral energy distribution (SED) using the flexible stellar population synthesis (FSPS, \cite{Conroy_09}) module. We use the {\tt Prospector} \cite{Johnson_21} software to run a Markov Chain Monte Carlo (MCMC) sampler \cite{Foreman-Mackey_13}. We assume an exponentially decaying star formation history (SFH), and a flat prior on the five free model parameters: stellar mass ($M_{\star}$), stellar metallicity ($Z$), $E(B-V)$ extinction (assuming the extinction law from\cite{Calzetti_2000}), the stellar population age ($t_{age}$) and the e-folding time of the exponential decay of the SFH ($\tau_{\rm{sfh}}$).

The observed and modeled SEDs are shown in EDF~7, alongside the best-fitted parameters and their 
uncertainties. Of a particular interest is the total stellar mass ($M_{*} \approx 2\times 10^{10}$ \msol) that can be used to 
obtain an order of magnitude estimate of \mbh using  \cite{Greene2020} $M_*$-\mbh relation, we obtain $\rm{log}$(\mbh/\msol)$ = 6.9 \pm 0.7$, where the uncertainty accounts for both the statistical and the spread of the scaling relation, but it is dominated by the former. The resulting \mbh value is consistent, within the error bar, with the one obtained from the $\sigma_{*}$-\mbh 
relation. We adopt the value obtained from $\sigma_{*}$, given the smaller spread and systematically more consistent values obtained by the $\sigma_{*}$-\mbh \cite{Kormendy2013,Greene2020}.

Using a bolometric correction \cite{Pennell} and the measured extinction corrected [O\,\textsc{iii}]$\lambda$5007\AA\xspace luminosity of $
\sim 5 \times  10^{38}$ \ergs we estimate the upper limit for bolometric luminosity of this LLAGN to be \Lbolquie$\leq 2 \times 10^{42}$ \ergs.
From the 2.0-10.0 keV luminosity upper limit ($\sim 9 \times 10^{40}$\ergs, see Estimating the XRT X-ray upper limit'') and \cite{Duras2020} bolometric correction we also estimated an 
upper limit of  \Lbolquie$\leq$ $9 \times 10^{41}$ \ergs. Given that the [O\,\textsc{iii}] derived value represents the mean accretion rate 
in the last few Myr (which represents the time to ionize the entire Narrow Line Region, NLR) and that star-formation likely contributes to a considerable fraction of 
its luminosity, while the hard X-ray (2.0-10.0 keV) is a more precise estimate of the current accretion state, we adopt $9 \times 10^{41}$ \ergs as 
our estimate of the upper limit for the bolometric luminosity of the LLAGN in \target's host.
Combined, the measured \Lbolquie and \mbh result in an Eddington ratio 
$\lambda_{\rm Edd} =\Lbol/\LEdd <$ 0.002, confirming the extremely low accretion rate of \target's host galaxy previous to the eruptions.

\section{Constraints on the start of the eruptions}\label{sec:constraint_begin}
We searched for archival X-ray observations at the \target's position in order to constrain the start of the eruptions: there is no previous X-ray detection at this position. The X-ray upper limit server \cite{up_lim_server} returns a 0.2-2.0 KeV flux $\leq 3.3 \times 10^{-13}$ from a 1990 $\sim$ 330s observation of ROSAT all-sky survey (RASS, \cite{Voges1999}), and $\leq 5.2 \times 10^{-13}$ \fluxunits from a 2005 $\sim$ 9s exposure of the \xmm Slew Survey \cite{Saxton2008}. The field containing the position of \target was observed by XRT between 1-11 December 2021 with a three-day cadence and between 24 December 2021 and 8 January 2022, also at a three-day cadence.
There were no detections during this early monitoring, and all the upper limits were below the flux level of the first detection.
The non-detections during this early XRT monitoring put a hard constraint on the start of the QPEs, it is unlikely that no eruption would be detected during high-cadence month-long monitoring if they had already started. In order to quantify this likelihood, we perform the following series of simulations:

\begin{itemize}
    \item We shifted the exactly known XRT light curve (shown in Fig.~1, MJD 59752-59983) to the epochs of the early XRT monitoring (shown in red in EDF 1), maintaining its cadence and gaps.
    \item We checked if at least one of the early time observations matched a detection of the simulated light curve, given a $\pm 0.5$ day range.
    \item Repeated the process 10000 times but randomly changed the relative shift between the simulated light curve and the early monitoring observations in each iteration, ensuring the epochs of simulated light curves included the epochs of the early monitoring.
    \item Checked how many of these 10000 simulations have at least one detection.
\end{itemize}

\noindent From the simulations, we found that 88\% of the time, we would have made at least one detection if the known light curve (series of eruptions) were present during the early monitoring; this means that the probability of not observing such eruptions in the early monitoring is only 12\%. However, we note that this is driven mainly by the large period without long-lived (several days long) eruptions between MJD 59880 and MJD 59940. If the simulated light curve had only the six first consecutive long-eruptions (from discovery up to MJD 59880), then the probability of making no detections during the early monitoring would drop to ~0.1\%. Therefore, we can conclude that the eruptions in \target have, most likely, started between the end of early monitoring campaign and the date of the first detection (i.e. between 8 January and 22 June 2022). In EDF 1, we show the long-term light curve with all the archived non-detections as a function of time.

\renewcommand{\figurename}{Extended Data Figure}
\renewcommand{\tablename}{Extended Data Table}
\setcounter{figure}{0}
\setcounter{table}{0}

\begin{figure*}[b!]
    \centering
    \includegraphics[width=0.8\textwidth]{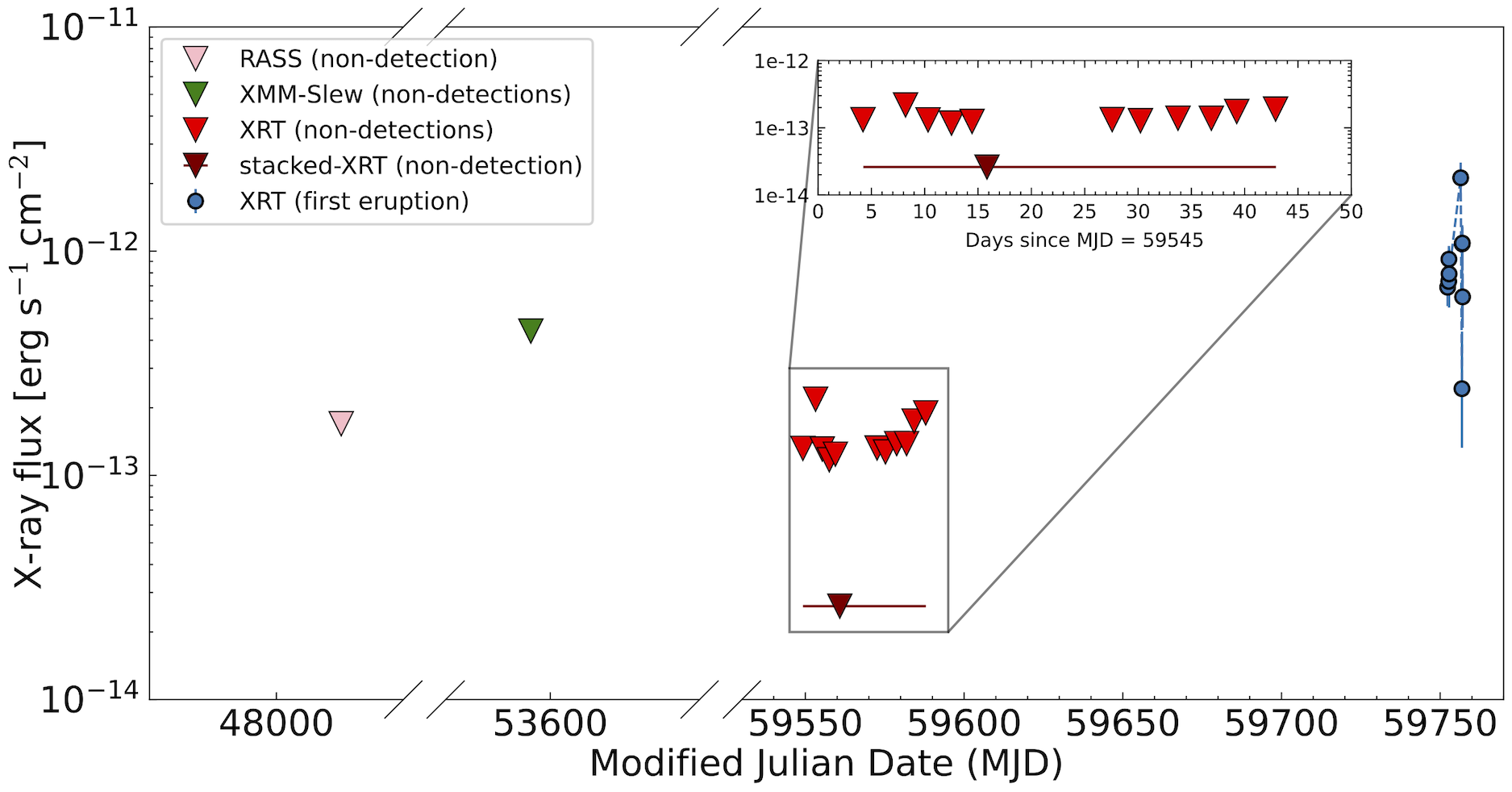}
    \caption{ \textbf{Constraint on the beginning of the eruptions in \target:} historical X-ray light curve, 3$\sigma$ upper limits from non-detection in 1990 by RASS (pink triangle), in 2005 by \xmm-Slew survey (green triangle) and multiple \swift/XRT observations between 1 December 2021 and 8 January 2022 (red triangles). The multiple consecutive non-detections of XRT constrain that the eruptions may have started between 8 January and 22 June 2022 --  date of the first detection by \swift/XRT (blue points).}
    \label{fig:constraint_xray}
\end{figure*}

\begin{figure*}[b!]
    \centering
    \includegraphics[width=0.90\textwidth]{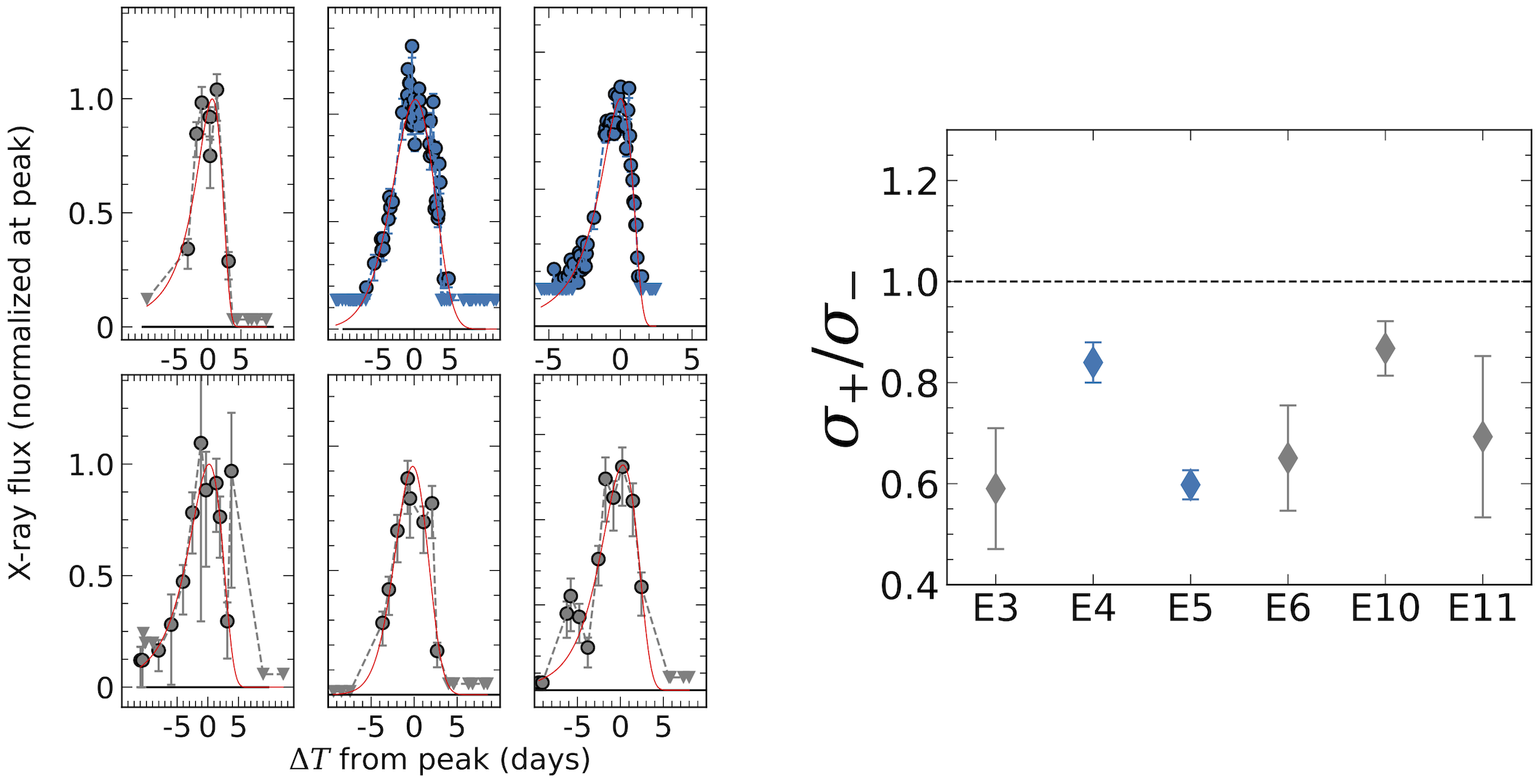}
    \caption{{\bf Eruptions shape fitting.}. {\bf Left}: Fit of asymmetric Gaussian profile to the six best-sampled eruptions: around epochs \textit{E3}, \textit{E4}, \textit{E5}, \textit{E6}, \textit{E10}, and \textit{E11}. {\bf Right}: ratio of $\sigma_{+}$ and $\sigma_{-}$ showing the slight asymmetric nature of \target's eruption. Error-bars represent 1$\sigma$ uncertainty. }
    \label{fig:Erup_shape}
\end{figure*}

\newpage

\begin{figure*}[h!]
    \centering
   \includegraphics[width=1.0\textwidth]{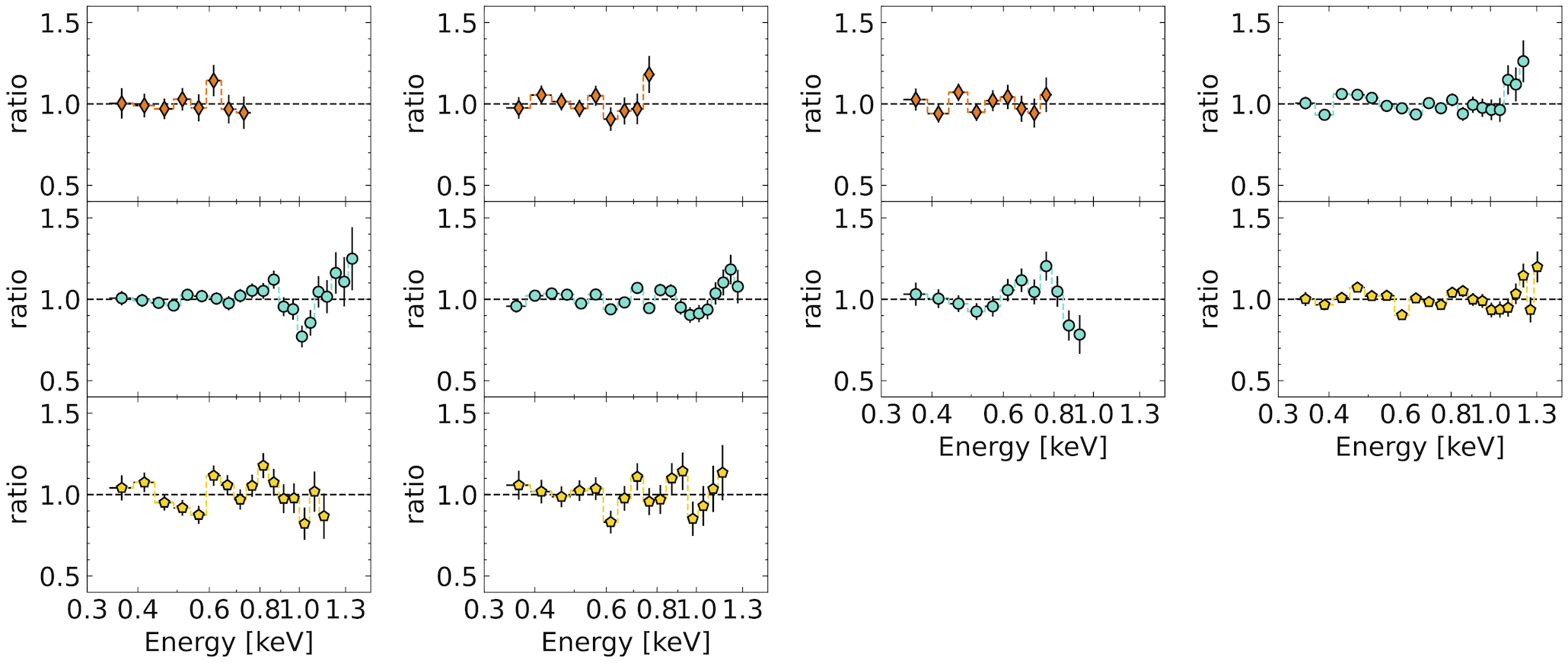}
    \caption{{\bf Residuals of the stacked spectral analyses.}. The order (left to right panels) represent distinct eruptions while the color and vertical panels represent distinct phases of each eruption: orange (rises), cyan (peaks) and gold (decays). The order and colors are the same as in Fig.~1. Error-bars represent 1$\sigma$ uncertainty.}
    \label{fig:resdiuals}
\end{figure*}

\begin{figure*}[b!]
    \centering
    \includegraphics[width=0.7\textwidth]{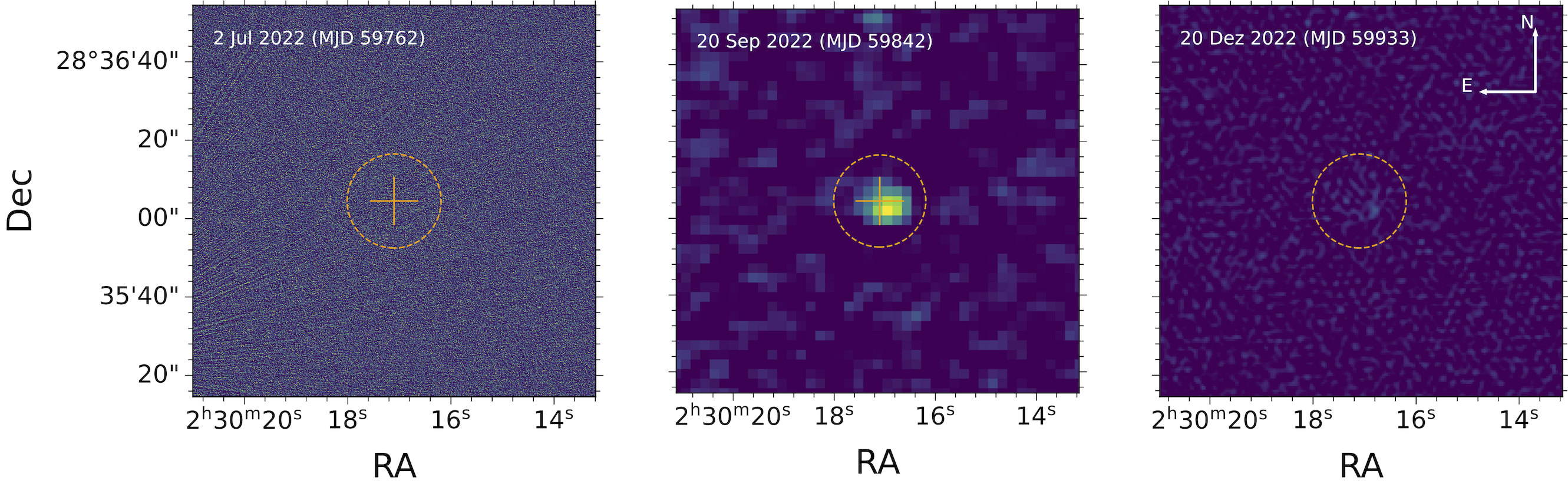}
    \caption{\textbf{Radio (VLA) images:} A transient radio source is detected in the second radio observation (middle panel) on MJD 59842 with a flux of  $93\pm7$ $\mu$Jy (13$\sigma$ detection). No source is detected in the first and third observations (left and right panels), with upper limits of 15 $\mu$Jy and 25 $\mu$Jy respectively. The orange cross marks the peak of the X-ray emission, and the orange circle the Kron radius (12\arcsec) of the host galaxy.}
    \label{fig:VLA}
\end{figure*}

\begin{figure*}[h!]
    \centering
    \includegraphics[width=0.7\textwidth]{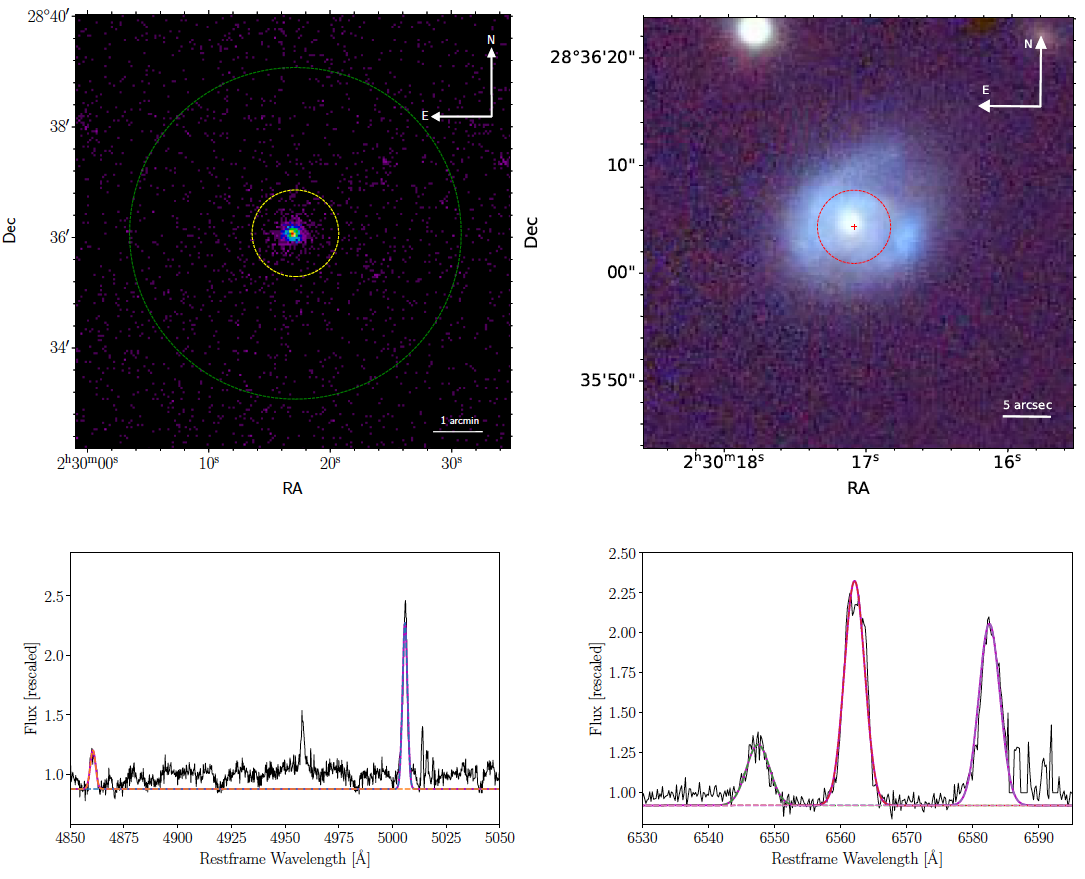}
  
    \caption{\textbf{\target position and host identification}. {\bf Top left}: \swift/XRT stacked images. Yellow 47\arcsec circle represents the 90\% region of the XRT point spread function, and was the radius used for extraction. Green circle is the \nicer FoV, no other source is present. {\bf Top Right}: Pan-STARRS \textit{i}/\textit{g}/\textit{r} bands composed image of \target's host galaxy. Red cross show the location of the peak of the XRT emission and red circle (radius = 3.4\arcsec) represents the 2.7$\sigma$ uncertainty on the position. The X-ray emission is consistent with the nucleus of the galaxy. {\bf Bottom}: Continuum normalized X-shooter optical  spectrum of the nuclear 1\arcsec of the host galaxy, in the H$\beta$+[O\,\textsc{iii}] (left) and H$\alpha$+[N\,\textsc{ii}] (right) regions.}
    \label{fig:host}
\end{figure*}

\begin{figure}[b!]
    \centering
    \includegraphics[width=0.9\textwidth]{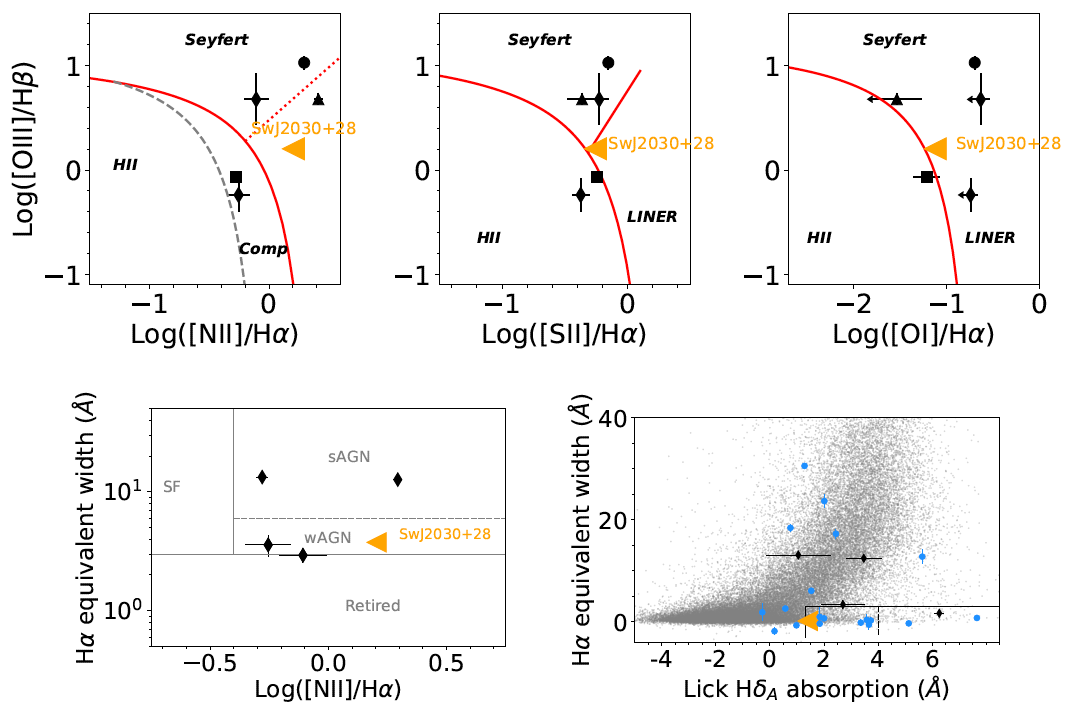}
    
    \caption{\textbf{Diagnostic diagrams of the host galaxy}. {\bf Top}: \target in the BPT diagnostic diagram, located above the \cite{Kewley2001} theoretical upper limit for star-formation ionization (red continuous line). Black diamonds represent the 4 known QPE hosts in all panels \cite{Wevers2022}. {\bf Middle}: \target in the WHAN diagnostic diagram, further showing that the nucleus likely hosts a weak AGN. {\bf Bottom}: the Lick H$\delta$ absorption index as a function of H$\alpha$ EW diagram. Grey points show SDSS galaxies for reference; blue circles represent TDE host galaxies. The black dash-delimited (solid) box indicates where QBS (E+A) galaxies are located. These galaxies make up 2.3\% and 0.2\% of the selected SDSS galaxies, respectively. Error-bars are 1$\sigma$ uncertainties in all panels.}
    \label{fig:diagdiag}
\end{figure}

\clearpage
\begin{figure}[h!]
    \centering
    \includegraphics[width=0.7\textwidth]{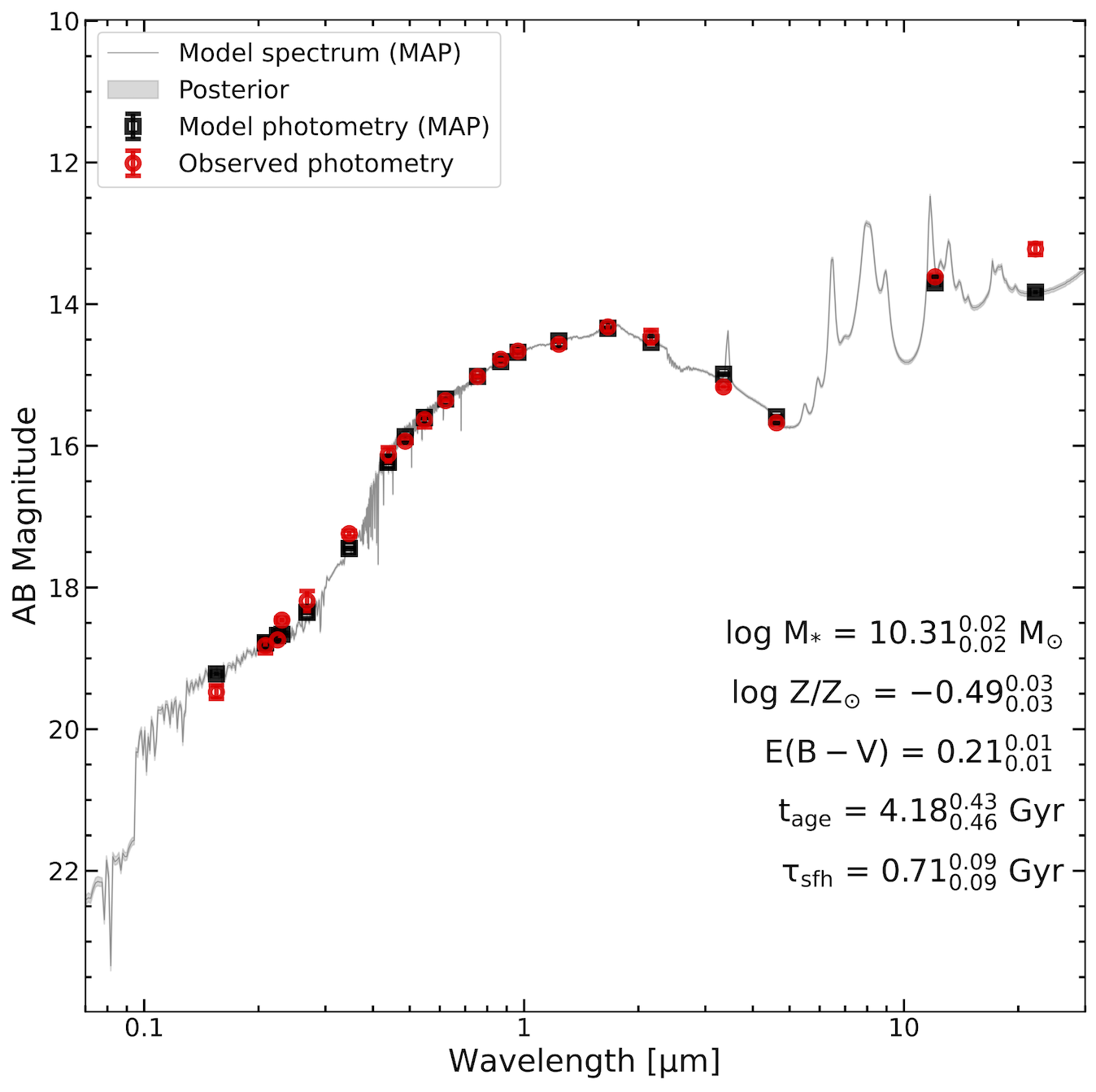}
    \caption{\textbf{Broad-band spectral energy distribution (SED) of host galaxy}. Red points show the observed archival photometry, black point the maximum a posteriori (MAP) best-fitted mode, and grey line the MAP best-fitted spectrum. Best-fitted parameters (see text for details) for the model are shown in the lower right. Error-bars are 1$\sigma$ uncertainties.}
    \label{fig:host_sed}
\end{figure}

 \clearpage
 
\begin{figure*}[]
    \centering
    \includegraphics[width=0.8\textwidth]{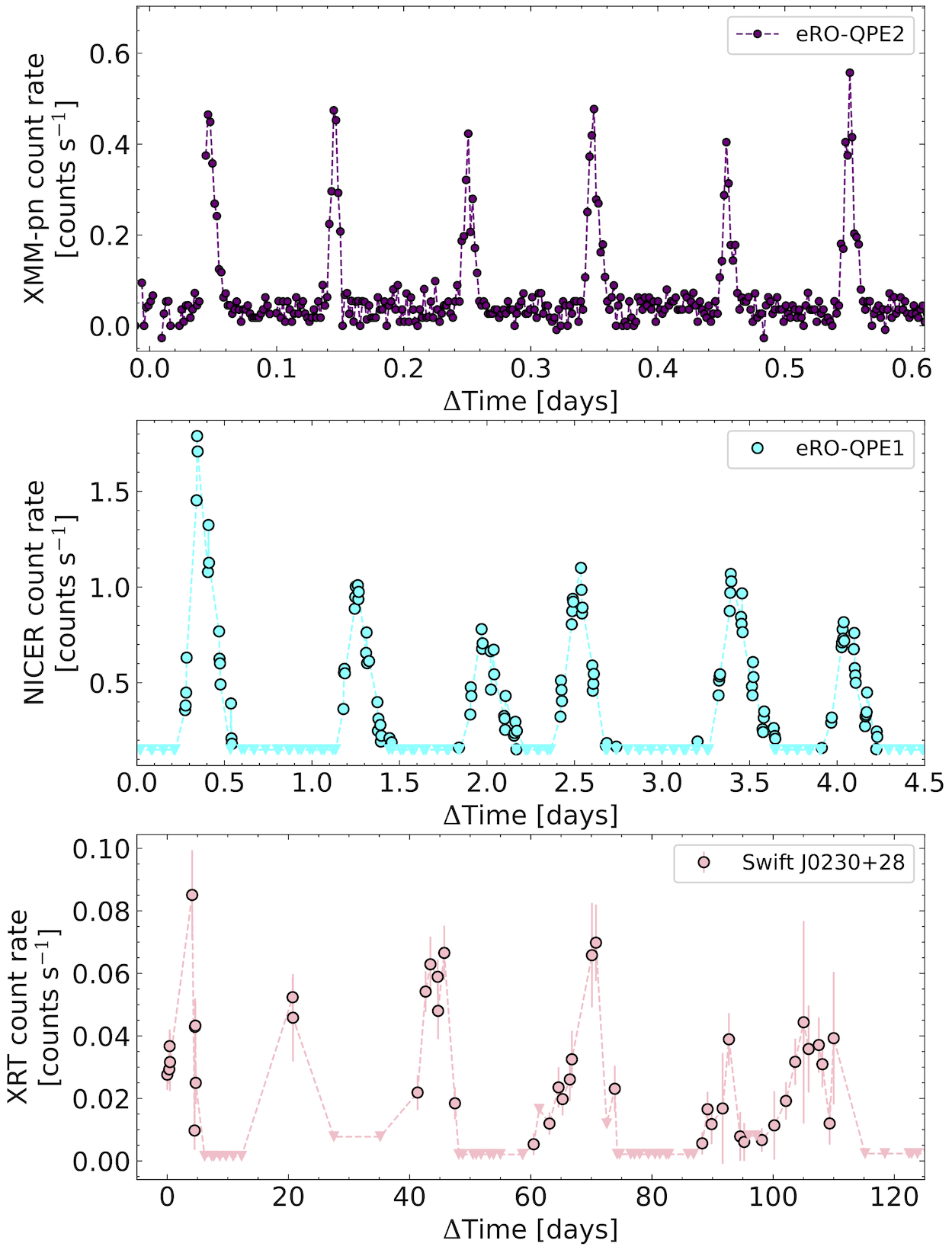}
    \caption{\textbf{Comparison of QPEs light curves.} {\bf Top}: \xmm-pn light curve of eRO-QPE2. {\bf Center}: \nicer light curve of eRO-QPE1. {\bf Bottom:} \swift/XRT light curves of \target. All three panels show six consecutive eruptions for each source, the distinct time scales are clearly given the x-axis range: 0.6 days for eRO-QPE2, 4.5 days for eRO-QPE1 and 120 days for \target. Error-bars are 1$\sigma$ uncertainties.}
    \label{fig:lc_comparison}
\end{figure*}

 \clearpage

\begin{table}[ht!]
    \centering
    \caption{{\bf Summary of time-resolved X-ray spectra analyses with absorbed thermal model on stacked spectra}. The stacked \nicer spectra are fit with {\tt tbabs*zTBabs*zashift(diskbb)} model using {\it XSPEC} \cite{xspec}. {\bf Start} and {\bf End} represent the start and end times (in units of MJD) of the interval used to stack the data. {\bf $\log L_{\rm 0.3-2.0 keV}$} is the logarithm of the integrated absorption-corrected luminosity in 0.3--2.0 keV range in units of \ergs, assuming the best-fitted model. $N_H$ is the best-fitted intrinsic column density in units of $10^{20}$ cm$^{-2}$. {\bf $T_{in}$} is the best-fitting inner disk temperature of \texttt{diskbb} in eV.  Uncertainties represent 1$\sigma$ level.}
    \ttabbox[\linewidth]{
    \begin{tabular}{ccccccc}
   
    {Phase}    &   {Start}   & {End} & ${\rm \ log \  L}_{\rm {0.3-2.0 keV}} $       & {$N_H$} & {$T_{in}$} & 
    {$\chi^2$/d.o.f}\\
     &  (MJD)  & (MJD)      & [\rm \ergs]          & [$10^{20}$ cm$^{-2}]$& [eV]      &    \\
    \hline
     Rise-E4   & 59814.2 & 59818   & 42.80 $\pm$ 0.11 & 3$^{+4}_{-3}$ & 124$^{+11}_{-9}$ & 5.1/6   \\
 Peak--E4  & 59819.3 & 59821.9 & 42.96 $\pm$ 0.03 & 2$^{+3}_{-1}$ & 177$^{+4}_{-4}$  & 28.5/19 \\
 Decay-E4  & 59823.1 & 59824.6 & 42.87 $\pm$ 0.05 & 3$^{+4}_{-3}$ & 172$^{+7}_{-7}$  & 24.7/14 \\
 Rise-E5   & 59841.9 & 59844.2 & 42.33 $\pm$ 0.06 & 1$^{+2}_{-1}$ & 140$^{+4}_{-12}$ & 6.4/6   \\
 Peak--E5  & 59844.4 & 59846.1 & 42.74 $\pm$ 0.02 & 1$^{+1}_{-1}$ & 202$^{+2}_{-6}$  & 29.6/17 \\
 Decay-E5  & 59846.1 & 59846.6 & 42.53 $\pm$ 0.02 & 1$^{+1}_{-1}$ & 209$^{+5}_{-7}$  & 14.4/13 \\
 Rise-E6   & 59850.7 & 59853.9 & 42.52 $\pm$ 0.09 & 2$^{+4}_{-1}$ & 131$^{+10}_{-8}$ & 5.9/6   \\
 Peak-E6   & 59855.8 & 59858.9 & 42.86 $\pm$ 0.07 & 3$^{+4}_{-1}$ & 149$^{+12}_{-4}$ & 17.2/9  \\
 Peak-E10  & 59949   & 59951.2 & 42.93 $\pm$ 0.03 & 1$^{+2}_{-1}$ & 165$^{+3}_{-3}$  & 25.7/16 \\
 Decay-E10 & 59951.4 & 59953.5 & 42.84 $\pm$ 0.02 & 1$^{+1}_{-1}$ & 214$^{+4}_{-4}$  & 27.9/18 \\
    \hline
    \end{tabular}
    }  
    
    \label{tab:stacked_spec}
\end{table}

\begin{table}[ht!]
    
    \footnotesize
    \ttabbox[]{
    \begin{tabular}{|c|c|c|c|c|c|c|}
    \hline
\multicolumn{2}{|c|}{Source}                                                                                            & \multicolumn{1}{c|}{eRO-QPE2}       & \multicolumn{1}{c|}{GSN 069}               & \multicolumn{1}{c|}{RX J1301}               & \multicolumn{1}{c|}{eRO-QPE1} & \target                                                                                 \\ \hline
\multicolumn{2}{|c|}{\begin{tabular}[c]{@{}c@{}}Mean recurrence \\ time\end{tabular}}                                   & \multicolumn{1}{c|}{2.4h}           & \multicolumn{1}{c|}{9h}                    & \multicolumn{1}{c|}{4.7h}                  & \multicolumn{1}{c|}{18.5h} & 22d                                                                                     \\ \hline
\multicolumn{1}{|c|}{\multirow{3}{*}{\begin{tabular}[c]{@{}c@{}}Deviation \\ from\\  periodicity\end{tabular}}} & Pattern  & \multicolumn{2}{c|}{`long-short'}                                                & \multicolumn{3}{c|}{irregular} \\ \cline{2-7} 
\multicolumn{1}{|c|}{}                                                                                       & Absolute & \multicolumn{1}{c|}{2.3h-2.7h}      & \multicolumn{1}{c|}{8.5h-9.5h}             & \multicolumn{1}{c|}{3.5h-5.5h}             & \multicolumn{1}{c|}{12h-24h$^{a}$} & 18d-25d$^b$                                                                                \\ \cline{2-7} 
\multicolumn{1}{|c|}{}                                                                                       & Fraction & \multicolumn{2}{c|}{$\sim \pm$ 10\%}                                               & \multicolumn{1}{c|}{$\sim \pm$ 20\%}         & \multicolumn{1}{c|}{30\%-50\%$^{a}$} & $\sim \pm$ 30\%$^{b}$                                                                        \\ \hline
\multicolumn{2}{|c|}{\begin{tabular}[c]{@{}c@{}}X-ray luminosity \\ at peak$^c$ (\ergs)\end{tabular}}      & \multicolumn{1}{c|}{$\sim 10^{42}$}   & \multicolumn{1}{c|}{$\sim 5 \times 10^{42}$} & \multicolumn{1}{c|}{$\sim 2 \times 10^{42}$} & \multicolumn{1}{c|}{$\sim 10^{43}$} & $\sim 6 \times 10^{42} $                                                                       \\ \hline
\multicolumn{2}{|c|}{\begin{tabular}[c]{@{}c@{}}Amplitude$^{c,d}$\ (count rate)\end{tabular}}                & \multicolumn{1}{c|}{$\sim$ 10}        & \multicolumn{1}{c|}{$\sim$ 20}               & \multicolumn{1}{c|}{$\sim$ 15}               & \multicolumn{1}{c|}{100-200} & $\geq$ 100                                                                                \\ \hline
\multicolumn{2}{|c|}{\begin{tabular}[c]{@{}c@{}}Temperature evolution\\ per eruption\end{tabular}}                      & \multicolumn{4}{c|}{cool $\rightarrow$ warm  $\rightarrow$ cool} & cool $\rightarrow$ warm                                                                 \\ \hline
\multicolumn{2}{|c|}{Eruption shape}                                                                                    & \multicolumn{1}{c|}{$\sim$ symmetric} & \multicolumn{1}{c|}{$\sim$ symmetric}        & \multicolumn{1}{c|}{$\sim$ symmetric}        & \multicolumn{1}{c|}{\begin{tabular}[c]{@{}c@{}}asymmetric \\ (longer decay\\  than rise)\end{tabular}} & \begin{tabular}[c]{@{}c@{}}slight asymmetric\\ (longer rise \\ than decay)\end{tabular} \\ \hline
\multicolumn{2}{|c|}{log \mbh (\msun)}                                                                                  & \multicolumn{1}{c|}{5.0 $\pm$ 0.5}    & \multicolumn{1}{c|}{6.0 $\pm$ 0.5}           & \multicolumn{1}{c|}{6.6 $\pm$ 0.4}           & \multicolumn{1}{c|}{5.8 $\pm $0.5}                                                                       & 6.6 $\pm$ 0.4                                                                             \\  \hline
    \end{tabular}
   
    \caption{\textbf{ Properties of \target as compared to quasi-periodic eruption sources (QPEs).} Notes: a) This is a conservative estimate, given that as shown by \cite{Arcodia2022} some of the eruptions overlap with each other, which means an even lower separation between two consecutive eruptions. b) This is a conservative estimate, based on the fitted peak (see `X-ray light curve' in Methods) of the well-sampled eruptions, given that some eruptions were not observed around the LSP peak period (see e.g., \textit{E9} in Fig. 1), which means an even larger separation between two consecutive eruptions. c) In the 0.3-2.0 keV band. d) Defined as the ratio between count rates of the peak and quiescent states. e) Based on whether the eruptions can be well fitted by a Gaussian, or some asymmetric function, e.g., asymmetric Gaussian or Gaussian-rise plus power-law decay, is necessary. }}
    
    \label{tab:QPE}
    
\end{table}

\newpage

{\bf Acknowledgments.}\\ 
During the refereeing process of this manuscript, Evans et al. 2023 
\cite{Evans2023}, published a paper presenting a focused investigation of \target. The author's data set does not include our \nicer and radio (VLA) data, but their science case and physical interpretation are similar to ours.

MG and SG are supported in part by NASA grants 80NSS23K0621
and 80NSSC22K0571. DRP was supported by NASA grant 80NSSC22K0961 for this work. DRP and RR acknowledge support from NASA grant 80NSSC19K1287. MZ was supported by the GACR Junior Star grant no. GM24-10599M.
TW warmly thanks the Space Telescope Science Institute for its hospitality during part of this work. PS has been supported by the fellowship Lumina Quaeruntur No.\ LQ100032102 of the Czech Academy of Sciences. This work was supported by the Ministry of Education, Youth and Sports of the Czech Republic through the e-INFRA CZ (ID:90140) and LM2023047 (VK). VW was supported by the Charles U. research programme PRIMUS 23/SCI/019. R.A acknowledges support by NASA through the NASA Einstein Fellowship grant No HF2-51499 awarded by the Space Telescope Science Institute, which is operated by the Association of Universities for Research in Astronomy, Inc., for NASA, under contract NAS5-26555. This work was supported by the Australian government through the Australian Research Council’s Discovery Projects funding scheme (DP200102471). ERC acknowledges support from the National Science Foundation through grant AST-2006684, and from the Oakridge Associated Universities through a Ralph E.~Powe Junior Faculty Enhancement award. ECF is supported by NASA under award number 80GSFC21M0002. KDA acknowledges support from the National Science Foundation under Grant No. AST-2307668.

The authors wish to recognize and acknowledge the very significant cultural role and reverence that the summit of Mauna Kea has always had within the indigenous Hawaiian community. We are most fortunate to have the opportunity to conduct observations from this mountain.
NICER is a 0.2-12 keV X-ray telescope operating on the International Space Station. The
NICER mission and portions of the NICER science team activities are funded by NASA. Based on observations made with ESO Telescopes at the La Silla Paranal Observatory under programme ID 110.2599. 
The National Radio Astronomy Observatory is a facility of the National Science Foundation operated under cooperative agreement by Associated Universities, Inc.

{\bf Author contributions:} 
MG led the overall project, wrote a large portion of the paper, performed part of the data analyses, and was the PI of the \nicer DDT proposals. DRP reduced the \nicer data and performed part of the X-ray analyses. MZ and ERC led the theoretical/modeling portion of the project and wrote parts of the paper. VJ, PS, and VK contributed to the modeling/theory portion of the paper. TW was the PI of the VLT DDT program, and wrote part of the paper. SVV, KDA (PI), and JMJ are the team leads of the VLA program; SVV performed the radio data reduction and analyses. RR, KG, and ECF performed the \nicer observations. SG, FT, YY, and RA contributed to the data gathering, analyses, and interpretation, as well as paper writing and discussion.

{\bf Competing interests:} The authors
declare that there are no competing interests.

{\bf Data and materials availability:} All the \nicer and \swift data presented here are public and can be found in the NASA archives at the following URL: \url{https://heasarc.gsfc.nasa.gov/cgi-bin/W3Browse/w3browse.pl}. The VLA data are available from the NRAO archives, at \url{https://data.nrao.edu/portal/#/}. X-shooter spectra will be avalible at ESO archive after the 12 months proprietary is passed. Keck/ESI will be shared under request to the corresponding authors. The GRMHD simulation described as part of the  ``Accretion disc – perturber interaction'' section in Supplementary Section, is available upon request to the corresponding author. The data underlying the multi-wavelength light curves presented in Fig.~1 are available at \url{https://zenodo.org/records/10238766}.
 \newpage


\clearpage

\renewcommand{\figurename}{Supplementary Information Figure}
\renewcommand{\tablename}{Supplementary Information Table}
\setcounter{figure}{0}

\section*{{\Huge Supplementary Information}}

\section*{Quasi-periodic erupters and Swift J0230+28}\label{sec:qpe_comparison}

In the last few years, quasi-periodic X-ray eruptions (QPEs) were discovered from the nuclei of some nearby galaxies \cite{Miniutti2019,Giustini2020,Arcodia2021}. Their emission is characterized by soft X-ray ($E < 2.0$ keV) recurrent eruptions with amplitudes in 0.3-2.0 keV count-rate as high as $\sim 200$ \cite{Arcodia2021,Arcodia2022} compared to their quiescent level. These QPE sources show no accompanying variability in the optical and UV bands, and their host galaxies show no clear signs of recent nuclear activity (e.g., hard X-ray emission or IR colors and variability), including the absence of optical broad emission lines in their optical spectra. Furthermore, all QPEs are hosted by black holes with masses $\leq 10^7$ \msol. Four QPE sources are now confirmed: GSN~069 \cite{Miniutti2019},  RX~J1301.9+2747 (hereafter RX~J1301,\cite{Giustini2020}), eRO-QPE1 and eRO-QPE2 \cite{Arcodia2021}. Here we do not include XMMSL1 J024916.6-041244 \cite{Chakraborty2021}, because only 1.5 eruptions were detected; hence it is only a QPE candidate. The mean recurrence time between eruptions varies from 2.4 hours ($\sim$ 9 ks) for eRO-QPE2 up to 18.5 hours ($\sim$ 70 ks) for eRO-QPE1. The duration of the eruptions varies between 0.3 hours (1.2 ks), for RX~J1301 and 7.5h (27ks), for eRO-QPE1. The amplitudes (in the 0.3-2.0 keV band) of the eruptions vary between $\sim$ 10 for eRO-QPE2 and 100-200 for eRO-QPE1. At their peak, their 0.3-2.0 keV luminosity varies between $\rm{few} \times 10^{42}$ \ergs to $\rm{few} \times 10^{43}$ \ergs.

Although their eruptions are recurrent and quasi-periodic, they show a significant deviation from a strictly periodic signal. GSN~069 and eRO-QPE2  exhibit an oscillating `long/short' behavior of the recurrence time between eruptions, where the long and short recurrence time vary by $\sim 10\%$ from the mean value, with the short/long recurrence time of 8.5h/9.5h for GSN~069 and 2.35h-2.75h for eRO-QPE2. In the case of RX J1301, it is still not clear whether such `long/short' alternating behavior is present; however, the deviation from the mean period is even larger with recurrence time as long as 5.5h (20ks), and as short as 3.5h (13ks) observed \cite{Giustini2020}, which implies a deviation from the mean recurrence time that can be as large as 25\%. In the case of the longest and brightest of the QPE sources, eRO-QPE1, the instability in the recurrence times is even more extreme\cite{Arcodia2022}; it can be as short as 12h and as long 24h, with chaotic variability in the recurrence times and no clear `long/short' alternating behavior pattern. This implies a deviation from the mean recurrence time that can be as large as 50\% (see, e.g., Extended Data Figure 8 and the bottom panel of Figure 1 in \cite{Arcodia2022}). 
This case is even more extreme, given that some of the eRO-QPE1's eruptions are not isolated, but instead show multiple overlapping eruptions with different amplitudes \cite{Arcodia2022}.
A common characteristic of all the QPEs is that the rise is harder than decay \cite{Arcodia2022,Miniutti2023}; the four sources show an increase in their temperature from about 50-70 eV up to 100-250 eV at the peak of the eruptions, which evolves back to the low temperatures during their decay, i.e, a cool $\rightarrow$ warm $\rightarrow$  cool temperature evolution in each eruption.

\target shares several defining characteristics of QPE sources: i) recurrent quasi-periodic soft X-ray eruptions (see `X-ray light curve' in Methods); ii) no UV/optical variability counterpart to the X-rays is observed (see `UV/optical and radio counterparts' in Methods); iii) it originates from the nucleus of a galaxy with no clear sign of strong multi-wavelength AGN emission (see `Host galaxy' in Methods); iv) it is hosted by a low mass black hole, $\leq 10^{6.6}$ \msol; v) the luminosities at the peak of the eruptions are between $\sim 10^{42}-10^{43}$ \ergs. 

The most obvious difference between \target and the four known QPE sources is the recurrence time. The peak-to-peak separation between the eruptions in \target varies between 18-25 days (1300-2160 ks), with a mean recurrence time of $\sim$22 days, which is a factor of $\sim$ 25 longer than the longest QPE source. However, it should be noted that all four known QPE sources were discovered either in the light curve of a long \xmm stare observation (within a single OBS ID, which are at most $\sim$120ks long \cite{Miniutti2019,Giustini2020}) or by consecutive 4h separated observations from one of \srge all-sky surveys \cite{Arcodia2021}. Both methods are biased to discover sources with recurrence times that are shorter than a $\rm{few}$ days. Therefore, it is not obvious that the longer time-scale of \target signals an intrinsic difference between \target and the known QPE population.

Unlike most of the other QPE sources, \target is not (yet, given the lack of \xmm observation) detected in the quiescent phase (between eruptions), and the derived upper limit on the quiescent luminosity leads to a lower limit on the amplitude to be $\geq 100$ for the strongest eruptions. Given the upper limit on the flux of the quiescent state, only \xmm may be able to detect it or derive a deeper limit; however, at the time of this publication \target was not yet observed by \xmm.

To highlight the similarities in behavior and differences in time-scales between previous known QPE sources and \target, in Extended Data Figure 8 we show the light curves containing six consecutive eruptions for the source with the shortest recurrence time (eRO-QPE2), the one with the longest recurrence time (eRO-QPE1), and \target. While six eruptions occurred in a period of less than one day in eRO-QPE2 and less than one week in eRO-QPE1, four months of observations were necessary to sample six eruptions in \target. 

Another distinction between \target and the four known QPE sources is the spectral evolution. \target is softer during the rise and harder during the decay, which runs contrary to the behavior of the four known QPE sources. The temperature evolution in each of \target's eruptions shows an increase from $\sim$100 eV at the rise to $\sim$200 eV at the decays, but with no decrease in the temperature in a single eruption (see `X-ray spectra' in Methods), i.e. a cool $\rightarrow$ warm temperature evolution in each eruption; while all the four known QPEs sources show a decreasing temperature during their decays. 

Another difference between known QPEs sources and \target is the shape of the eruptions. While QPE sources show either a quasi-symmetric eruption \cite{Giustini2020} or a faster rise and a slower decay \cite{Miniutti2019,Arcodia2021}, \target shows the opposite behavior: the rise phase is $\sim 30\%$ longer than the decay phase (see `X-ray light curves' in Methods).

To date, no systematic and simultaneous radio and X-ray monitoring campaign of QPE sources was performed, however GSN\,069 \cite{Miniutti2019} appears to show weak but steady radio emission. The same appears to be the case for eRO-QPE2, while eRO-QPE1 is currently undetected in radio (Arcodia et al., in prep.). RX~J1301, however, shows radio emission that is variable by more than a factor of 2.5 over a few days \cite{Yang+2022}. Radio variability of the same amplitude as in \target -- transient and by a factor of  at least 6  -- was not yet observed for the known QPE sources. However, the lack of extensive homogeneous monitoring and the different time scales involved do not allow us to confirm whether the radio variability is an intrinsic difference between \target and the QPE population.
In summary, although some differences in the spectral evolution and eruption shape are present, \target shows most of the defining characteristics of QPE sources; it shares many more similarities than differences with QPE sources, as well as much more similarities with QPE sources than any other known class of transient sources, including the repeating nuclear transient (RNTs) or also called repeating partial TDEs (repeating pTDEs), as we will discuss in `Repeating Nuclear Transients and Swift J0230+28'. In Extended Data Table 2 we compare the properties of \target with those of individuals QPE sources.
If we consider \target to be a QPE source, some tentative correlations seem to be present in the overall population, as shown by Extended Data Figure 8. The recurrence time and the duration of the eruptions seem to correlate positively, with a duty-cycle (duration/recurrence time) equal to $0.24 \pm 0.13$. A positive correlation also appears to be present between the recurrence time and amplitude of the eruptions. However, the mean recurrence time does not seem to be correlated with black hole mass.

\section*{Repeating Nuclear Transients and Swift J0230+28}\label{sec:rnt_comparison}

A distinct category of nuclear transients exhibiting a non-terminal evolution are known  as repeating nuclear transients (RNTs), alternatively referred to as repeating partial tidal disruption events (Repeating pTDEs) based on the leading theoretical interpretation \cite{Payne2021,Wevers2023,Liu2023}. Three clear cases of RNTs are known: ASASSN-14ko \cite{Payne2021}, AT2018fyk \cite{Wevers2023}, and eRASSt J045650.3-203750 (hereafter eRA J0456-20, \cite{Liu2023}). These transients exhibit recurrent outbursts that each individually resembles a TDE, occurring at varying intervals of $\sim$ 114 days up to 1200 days. All three sources display strong UV/optical emission ($L_{BB} \gg 10^{44}$ \ergs), and their X-ray spectra are distinct from the soft spectra. Furthermore, their individual outbursts are characterized by significant asymmetry, with decay times much longer than their rise, typical of all X-ray and optically selected TDEs \cite{Saxton2020,Hammerstein2023}. The black hole masses of these RNTs lie in the range of $7 \leq \rm{log(M_{BH}/M_{\odot})} \leq 7.7$.

ASASSN-14ko, initially discovered at optical wavelengths, exhibits recurring TDE-like UV/optical flares separated by an average period of approximately 114 days, with a peak integrated UV/optical luminosity of $\sim \rm{few} \times 10^{44}$ \ergs. These repeating flares have been observed since at least 2014 with minor variations in the observed period \cite{Payne2021,Payne2022,Payne2023}. The X-ray flares are much fainter at their maximum, reaching $10^{43}$ \ergs, while their periodicity differs from that of the optical/UV flares. Moreover, they seem to be delayed in relation to the optical/UV peaks. The X-ray spectra exhibit both a thermal and a coronal component.

AT2018fyk, another optically discovered nuclear transient \cite{Wevers2019_18fyk}, was initially identified as a regular TDE with a peak $L_{BB}$ of approximately $10^{45}$ \ergs and a peak 0.3-10 keV X-ray luminosity of a $\rm{few} \times 10^{43}$ \ergs. Initially, it displayed a thermal X-ray spectrum, but around 200 days after the optical peak, it underwent a spectral transition to a corona-dominated state \cite{Wevers2021}. The transient followed the expected decay characteristic of a TDE and was not detected by \xmm and \textit{Chandra} around 700 days after the peak. However, at approximately 1200 days after the first peak, it exhibited a second TDE-like outburst with a dramatic brightening in both UV/optical and X-rays \cite{Wevers2023}.

eRA J0456-20, discovered by \srge as an X-ray nuclear transient 
accompanied by UV emission, exhibits a repeating pattern at approximately 220-day intervals. The 0.3-10.0 keV luminosity during its first flare was a $\rm{few} \times 10^{44}$ \ergs, but it displayed a much stronger variation in both the period and luminosity compared to ASASSN-14ko. Similar to the other RNTs, its X-ray spectra comprise both thermal and corona components \cite{Liu2023}.

The only shared feature between RNTs and \target is their repetitive phases of activity followed by periods of quiescence. However, they differ significantly in every other aspects, namely: i) RNTs exhibit much brighter 0.3-10.0 keV X-ray luminosities ($\gg 10^{43}$ \ergs) compared to \target and most QPEs at peak; ii) While RNTs show more prominent brightness in the UV/optical range than in X-rays, \target does not exhibit any detected UV/optical emission; iii) RNT's s demonstrate highly asymmetric time evolution (except for eRA J0456-20), resembling TDEs with rapid rises and much slower decays, while \target's eruptions have a slightly slower rise than decay; iv) Unlike \target and most QPEs, RNTs do not display a purely soft/thermal X-ray spectrum.

The much higher luminosities of RNTs, both in the UV/optical bands and in the X-rays, as compared to \target, has also a great impact on possible physical interpretations, given that, assuming that the periods of high activity are accretion-driven, the large difference in luminosity means a large difference in the accreted mass per eruption/outburst/flare. While in RNTs a standard 0.1 efficiency on the conversion between mass and luminosity results in a considerable fraction of a solar mass ($\geq 10^{-1}$ \msol) accreted in each RNT's outburst \cite{Wevers2023}, an estimated accreted mass is only $10^{-5}-10^{-4}$ \msol per eruption in \target. This favors a repeating pTDE interpretation for RNTs and disfavors such a model for \target, as will be discussed in length below.

\section*{Physical models for Swift J0230+28}

Since the discovery of the first QPE source \cite{Miniutti2019} and the first RNT \cite{Payne2021} several models have been proposed to explain these phenomena. These models can be roughly divided into two classes: those involving accretion-disc instabilities and those involving bodies orbiting a massive black hole (i.e., extreme mass ratio inspirals; EMRIs). For the EMRI-related models a large diversity of possibilities for the nature of the orbiting body and the mechanism responsible emission have been presented. In the following sections we will explore each of the proposed (class of) models and their strengths and weaknesses as an explanation for \target's observed properties.

\label{sec:models}
\subsection*{Accretion disc instabilities}

The analysis of the alpha-disc theory predicts regions of thermal and viscous instabilities in the configuration space of accretion disks\cite{SS73,LightEard74,Hameury20} (it is not clear if these appear in first-principle simulations of turbulent disks, see \cite{DavTchekh20}), typically parameterized by the ratio of accretion rate with respect to the Eddington rate $\dot{m}$, black hole mass \mbh and viscosity parameter $\alpha$. The so-called ionization instability\cite{MeyMeyH81} occurs in the outer regions of the accretion disc with temperature $\sim 6000 K$, or typical radiation wavelengths $\sim 2500\,\rm nm$. Since we observe no variability at these wavelengths, we rule out this instability in \target. 

The case is more subtle for the so-called radiation-pressure instability\cite{LightEard74,ShakSun76}. This can be triggered for $\lambda_{\rm Edd} \gtrsim 0.1$ \cite{2020A&A...641A.167S} and manifests itself as an overheating and overfilling of the inner edge of the accretion disc that propagates outwards and leads to a drop in density and temperature of the accretion disc until it becomes extinct at some $R_{\rm ext}$. Consequently, the disc is gradually refilled by the external accretion rate until the instability is triggered again. The bolometric luminosity curve has two essential components, the slow rise of the luminosity and temperature of the refilling disk, and a peak or a set of close peaks corresponding to the triggering of the instability \cite{Janiuk2002,Merloni2006}. Both the rise and decay of the instability peaks are roughly symmetric with the width given by the viscous time at some characteristic radius where the instability is triggered, $t_{\rm v.trig.} \sim (H_{\rm trig.}/R_{\rm trig.})^{-2} (G M_{\rm BH}/R_{\rm trig.}^3)^{-1/2} \alpha^{-1}$, where $H_{\rm trig.}$ is the height of the disc in the unstable region before the instability is triggered. On the other hand, the refilling timescale (the duty cycle) is governed by the viscous time at the extinction radius $t_{\rm v.ext.} \sim (H_{\rm ext.}/R_{\rm ext.})^{-2} (G M_{\rm BH}/R_{\rm ext.}^3)^{-1/2} \alpha^{-1}$, where $H_{\rm ext.}$ is the height of the disc at the extinction radius.

The radiation-pressure instability scenario was used to model the QPEs in GSN 069 \cite{2020A&A...641A.167S,2022arXiv220410067S,2022ApJ...928L..18P}. The models generally predicted a much longer cycle than the $<1$ day QPEs, even though \cite{2022ApJ...928L..18P} were able to tune a number of free parameters of the disc to obtain correct light curves. Another semi-analytical disk-instability model aimed at the QPE phenomenon was recently proposed in \cite{Kaur2023}. There the support by a strong toroidal magnetic field corresponding approximately to the saturation point of the magneto-rotational instability was invoked to push the radiation-pressure instability towards the inner edge of the disk and shorter time-scales. For high magnetization values, the authors were also able to roughly reproduce the phenomenology of GSN 069 and other QPEs. Generally, all of these models require either a thin radiatively efficient disk or a combination of an outer thin disk and an inner ADAF \cite{2022arXiv220410067S}.

In \target the timescale is longer, which makes the case of these models somewhat more favorable. However, the lack of correlated UV/optical variability in \target is difficult to reconcile with the radiation-pressure-instability scenario, since one would generally expect the X-rays to be reprocessed by the outer parts of the disk. At the very least it suggests that the instability should be confined to the very inner parts of the accretion disc and that $R_{\rm trig.} \sim R_{\rm ext.} \sim R_{\rm ISCO} \sim 10^0-10^1\, R_{\rm g}$ in this scenario. As a result, we can compare the rise/decay time in \target $t_{\rm r/d} \sim t_{\rm v.trig.}  \sim 3$ days with the typical recovery time between the first 6 peaks $t_{\rm rec.} \sim t_{\rm v.ext.} \sim 15$ days. This would suggest $ (H_{\rm trig.}/R_{\rm trig.}) \sim 2.24 (H_{\rm ext.}/R_{\rm ext.})$ (assuming a constant $\alpha$). This is in accord with standard accretion disc theory, which suggests the unstable parts of the disc are hotter and geometrically thicker. Exploring the assumption that the instability is triggered only near the inner edge $R_{\rm trig.} \sim 10 R_{\rm g}$ and $\alpha \sim 10^{-2}$ leads to a prediction of 
\begin{align}
\frac{H_{\rm trig.}}{R_{\rm trig.}} \sim \left( \frac{G M_{\rm BH}}{R_{\rm trig.}^3} \right)^{-1/4}\! \alpha^{-1/2}\, t_{\rm v.trig.}^{-1/2}  \sim 0.25 \left(\frac{M_{\rm BH}}{10^6 M_{\odot}}\right)^{1/2} \left(\frac{\alpha}{0.01}\right)^{-1/2} \,.
\end{align} 
The prediction of standard accretion theory in the radiation-pressure dominated region and $R = 10 R_{\rm g}$ is that $(H/R) = 4 \times 10^{-1} \lambda_{\rm Edd}$. This means that the time-scales of \target are in principle consistent with a radiation-pressure instability triggered at $\sim 10 R_{\rm g}$ and $\lambda_{\rm Edd}\sim 0.6$. 
Nevertheless, we cannot reconcile the luminosity $\lambda_{\rm Edd} \sim 0.6$ with the strong constraint of \target's host emission before the beginning of the eruptions as well as between the eruptions. In general, for the unstable, radiation-pressure dominated inner disk to exist, the relative accretion-rate should be larger than $\dot{m}\gtrsim 0.002 \alpha_{0.1}^{-1/8} m_{6.6}^{-1/8}$ \cite{2008bhad.book.....K}, where $\alpha_{0.1}$ and $m_{6.6}$ are the viscosity parameter and the MBH mass scaled to 0.1 and $10^{6.6}\,M_{\odot}$, respectively. This is already in tension with the Eddington ratio upper limit of 0.002 before the eruptions. For the radiation-pressure instability to operate, the relative accretion rate needs to be even larger, $\dot{m}\gtrsim 0.15 \alpha_{0.1}^{41/29} m_{6.6}^{-1/29}$ \cite{2020A&A...641A.167S}, which is not met even during the eruptions.

The relatively long time-scale of the eruptions suggests that the phenomenon may also be related to Lense-Thirring precession and the warping of an accretion disc misaligned with the black-hole spin \cite{BarPett75}. Recent magneto-hydrodynamics simulations by \cite{Musoke2023} and  \cite{Liska2022} suggest that the warping creates an inner sub-disk that cyclically tears off and depletes, causing a burst on the time-scale $T_{\rm burst} \sim 1000 - 3000 R_{\rm g}/c$ and a duty cycle of $T_{\rm cyc.} \sim 10 000 - 50 000 R_{\rm g}/c$. These time-scales are an order or two too short to be consistent with the variability of \target and the mass range $M_{\rm BH} \sim 10^5 - 10^7 M_{\odot}$. Furthermore, the few broad irregular peaks seen in the limited number of simulations so far do not seem to correspond well to the features of \target. In addition, the disk-tearing instability is associated with low-viscosity thin disks that are highly inclined ($\gtrsim 45^{\circ}$) with respect to the equatorial plane \cite{2012MNRAS.421.1201N}, which is in tension with low accretion rates associated with \target that are associated with geometrically thick advection-dominated accretion flows. The classical Lense-Thirring precession and nutation of accretion flows and jets (see e.g. \cite{2023ApJ...951..106B}) should also reveal itself by rather mild and symmetric eruption due to the changing viewing angle ($\propto \cos{\iota}$), again being in tension with sharp and asymmetric eruptions of \target.

\subsection*{Extreme mass ratio inspirals (EMRIs)}

\subsubsection*{Repeating partial tidal disruption event}
One proposed explanation for the origin of quasi-periodic eruptions (QPEs) is a white dwarf orbiting about a massive black hole, such that the pericenter distance of the star coincides with its partial tidal disruption radius -- where a fraction of the envelope mass is removed by tides \cite{King2020} (see also \cite{Zalamea2010} in the context of extreme mass ratio inspirals, and \cite{macleod14} in the context of high-energy transients). The stripped mass then accretes onto the MBH, fueling the episodic eruptions.

In the context of \target, this model has a number of difficulties explaining the observations. For one, the mass of the black hole in the host galaxy has been inferred to be on the order of $\sim 10^{6}M_{\odot}$. Adopting a white dwarf mass of $0.6M_{\odot}$ and a corresponding radius of $R_{\star} = 0.011R_{\odot}$ \cite{Nauenberg1972} yields a tidal disruption radius of $0.6 GM_{\rm BH}/c^2$, i.e., close to a factor of 10 smaller than the direct capture radius of a non-spinning black hole. As noted by \cite{Cufari2022} (see also Table 1 of \cite{King2022} for specific values in the context of known QPE systems), the partial tidal disruption radius is a factor of $\sim 2$ larger for a 5/3-polytropic star \cite{Guillochon2013, Mainetti2017}, which is a good approximation for the density profile of a low-mass white dwarf. Relativistic effects \cite{Beloborodov1992} and the induced stellar rotation \cite{Golightly2019} also can both serve to increase the tidal disruption radius, but it is difficult to see how such a system could produce an observable signature (i.e., the white dwarf partially disrupted outside the horizon of the black hole; e.g., \cite{Zalamea2010}) unless the MBH mass is substantially smaller (e.g., \cite{king23} in the context of QPEs and the mass constraint on the black hole) than the one that follows from the $M_{\rm BH}$-$\sigma_{*}$ relation.

Second, if the X-ray eruptions arise from accretion and the accretion efficiency is of the order $10\%$, the mass accreted per eruption is $\sim 10^{-4} - 10^{-5}M_{\odot}$ (see `Eruptions Energetics' in Methods), implying that the mass lost by the star per orbit is a very small fraction of the total stellar mass. This suggests that the pericenter distance of the star is extremely fine tuned to coincide with the partial tidal disruption radius (there is a steep gradient in the amount of mass lost as a function of the pericenter distance near the partial disruption radius; \cite{Guillochon2013}), and begs the question of how the star achieved this necessary distance, because tidal dissipation and gravitational-wave emission effectively do not change the pericenter distance if the orbit is highly eccentric. A near-unity eccentricity is necessary in this case because the pericenter distance must be comparable to the horizon distance of the black hole (see the preceding paragraph), where the orbital time is comparable to $GM_{\rm BH}/c^3 \simeq few$ seconds, while the recurrence time is on the order of days.

Third, the return time of the debris stripped from a star in a tidal disruption event is $\sim T_{\star}\left(M_{\rm BH}/M_{\star}\right)^{1/2}$ \cite{Lacy1982, Rees1988}, where $T_{\star} = R_{\star}^{3/2}/\sqrt{GM_{\star}}$ is the dynamical time of the star, which is $\sim 2000$ seconds for a $0.6M_{\odot}$ disrupted by a $10^6M_{\odot}$ MBH (assuming such a scenario is possible, given the first shortcoming of this model noted above). While this timescale is in good agreement with the $\sim$ hour-long duration of QPEs, it is a factor of $\sim 100$ smaller than the $\sim$ few-day duration of most of the eruptions observed in \target. While the return time of the material stripped from the star in a partial TDE can be a factor of a few longer than the canonical timescale quoted above \cite{Guillochon2013}, it is difficult to see how one could achieve this large of a discrepancy in timescales with this model.

Fourth, there is the question of the stability of the mass transfer and the orbit itself in a white dwarf-SMBH system. As noted by \cite{Zalamea2010} and as discussed above, the highly relativistic nature of the encounter between the white dwarf and the SMBH means that, on orbital grounds alone, the black hole must be of lower mass and/or rapidly rotating (and prograde in the sense of the orbital angular momentum of the star) to ensure that the white dwarf is outside of the direct capture radius. Additionally, in order to explain the fairly regular amplitude of the eruptions in \target, one would require the mass transfer process to be stable, i.e., to not result in a rapid runaway in which more mass is stripped on each subsequent pericenter passage. A number of authors (e.g., \cite{Zalamea2010, metzger22, Linial2023, lu23}) have noted that this is particularly difficult to achieve in the case of a white dwarf star because of the polytropic mass-radius relationship $R \propto M^{-1/3}$, i.e., a reduction in the mass of the star results in an increase in its radius, rendering the star more susceptible to tidal stripping. Why this runaway process -- with stronger outbursts occurring with time as more mass is stripped on each subsequent pericenter passage -- is not occurring in \target is difficult to explain in the context of this model (as pointed out by \cite{lu23}, the high eccentricity required for such a scenario implies that the star is moving at a super-Keplerian speed near pericenter where the mass loss is taking place, which inhibits the otherwise-stabilizing influence of the transfer of angular momentum between the stellar orbit and the gaseous disc generated through the tidal stripping; however, this could be alleviated if the disc itself is also elliptical \cite{king23b}). 

Fifth (and final), although the eruptions in \target are quasi-periodic (see `X-ray light curve'), some of the high-activity periods were much fainter or much shorter in duration, or both, see, for example, epochs \textit{E7} to \textit{E9} in Fig. 1, but return to have its average behavior later, see epochs \textit{E10} to \textit{E11} in Fig. 1. This means that the mechanism responsible for the X-ray emission must have some associated instability/irregularity. While there are distinct timescales in a repeating partial TDE scenario, such as the orbital time of the star vs.~the fallback time of the debris (which are, in some cases, able to be individually determined from observations; \cite{Wevers2023}), it is difficult to see what would provide such `eruptions holidays' where the eruptions are either much fainter or much shorter.

A similar interpretation is that \target is powered by the repeated partial disruption of a star, but in this case, the orbiter is a main sequence star, as was suggested by \cite{Payne2021, Wevers2023, Liu2023} for the sources ASASSN-14ko, AT2018fyk, and eRASSt J0456-20, respectively. If the star is solar-like and of relatively low mass ($\lesssim 0.4 M_{\odot}$), then the recurrence time of $\sim 22$ days is in good agreement with the canonical fallback time from a TDE if the MBH mass is $\sim 10^{6.6}M_{\odot}$, and the similarity between the fallback time and the orbital time would likely yield enhanced variability in the properties of the eruptions. With an MBH mass of $10^{6.6}M_{\odot}$, a star with radius $R_{\star} = 0.4R_{\odot}$ and mass of $0.4 M_{\odot}$, the tidal radius would be $\sim 10 GM_{\rm BH}/c^2$, and the pericenter would therefore be relativistic (potentially explaining the lack of UV/optical emission). However, since the amount of stripped mass would be $\sim 10^{-3} - 10^{-4}$ of the stellar mass to explain the accretion-induced emission, the pericenter distance would have to be very fine-tuned. Similarly to the white dwarf case, those `eruption holidays' are also a problem for repeated partial disruption of a main sequence star.

\subsubsection*{Accretion disc -- perturber interaction}

Here we investigate the possibility that the series of X-ray eruptions are due to star-disc interactions, in which passages of inclined stars through the accretion flow induce density waves and shocks that modulate the inflow and the outflow rate (see e.g. \cite{2021ApJ...917...43S} for a detailed analysis).
Our constraint for the past activity of \target\ (see ``The host galaxy'' in Methods), i.e. $\lambda_{\rm Edd} <$ 0.002, leads to the fact that MBH is fed from a hot and diluted advection-dominated accretion flow (ADAF). During the eruptions, the X-ray luminosity reaches $L_{\rm peak}\sim 6.5\times 10^{42}\,{\rm erg\,s^{-1}}$, which corresponds to the Eddington ratio $\lambda_{\rm Edd}\gtrsim 0.02$ when the thermal, optically thick disc can in principle exist. Hence, \target could transition from the low/hard state to the high/soft state on the timescale of 18-25 days and the soft X-ray spectrum is detected for $5-9$ days, which corresponds to the  duration.

The quasi-periodic accretion rate outbursts of at least two orders of magnitude were reported in GRMHD simulations of \cite{2021ApJ...917...43S}, where the outburst recurrence timescale was clearly modulated by the orbital timescale of an orbiting stellar or black-hole perturber. In particular, for a nearly circular orbit, it is expected that the outbursts occur twice per orbital period, $P_{\rm orb}=2P_{\rm eruption}$, since the perturber crosses the disc twice for a non-zero inclination with respect to the disc midplane. \cite{2021ApJ...917...43S} simulated the perturber characterized by its influence radius which defines the interaction cross-section with the background hot ADAF flow.

When we consider $P_{\rm eruption}\sim 22$ days, then the orbital period of a perturber on a nearly circular orbit is $P_{\rm orb}\sim 44$ days, which results in the semi-major axis of 
\begin{equation}
a_{\star}/R_{\rm g}\simeq 985(P_{\rm orb}/44\,\text{days})^{2/3}(M_{\rm BH}/10^{6.6}\,M_{\odot})^{-2/3}\,,
\label{eq_semimajor_axis}
\end{equation}
which is too far to produce order-of-magnitude changes in the accretion rate 
given that the orbital velocity is $v_{\star}\simeq c/\sqrt{(r/R_{\rm g})}\sim 0.03c$.

When we assume that the stellar perturber is instead on an eccentric orbit and the X-ray eruptions 
is produced and detected during its pericenter passage, the orbital period is $P_{\rm orb}\sim 22$ days, which according to Eq.~\eqref{eq_semimajor_axis} results in $a_{\star}\sim 621\,R_{\rm g}$. To generate a density perturbation propagating at $\sim 0.1c$, the stellar pericenter velocity should be of a comparable order of magnitude, i.e. $v_{\rm \star,per}\sim 0.1\,{\rm c}$, which implies $e_{\star}\gtrsim 0.73$ or the pericenter distance should be $R_{\rm \star,per}\lesssim 167.7\,R_{\rm g}$. To ensure that the star can complete several $\sim$22-day orbits, it cannot be tidally disrupted, hence a Solar-type star must have the pericenter distance larger than the tidal disruption radius,
\begin{align}
    R_{\rm \star,per}>R_{\rm T} \sim R_{\star}\left(\frac{M_{\rm BH}}{m_{\star}}\right)^{1/3}
    \sim 18.8 \left(\frac{R_{\star}}{1\,R_{\odot}}\right) \left(\frac{m_{\star}}{1\,M_{\odot}}\right)^{-1/3} \left(\frac{M_{\rm BH}}{10^{6.6}\,M_{\odot}}\right)^{-2/3}\,R_{\rm g}\,,
\end{align}
i.e. its eccentricity must be smaller than $e_{\star}\lesssim 0.97$. Hydrodynamic processes (stellar winds) or the sufficiently massive perturber (a stellar-mass or intermediate-mass black hole) can ensure that the influence radius of an orbiting body is large enough to perturb the hot diluted flow. For an ADAF, this is achievable for the influence radius $\mathcal{R}\gtrsim 1\,R_{\rm g}$ for the pericenter distances $R_{\rm \star, per}<100\,R_{\rm g}$, see \cite{2021ApJ...917...43S}.

To support the feasibility of this scenario, we have simulated the transit of the star through the ADAF in the same way as in \cite{2021ApJ...917...43S}. The details of the numerical scheme are provided in \cite{2021ApJ...917...43S}.  We have used the modified version of the {\tt HARMPI} code \cite{2006ApJ...641..626N,2015MNRAS.454.1848R,2007MNRAS.379..469T}. The transit of the perturber through the flow is modeled via its influence radius $\mathcal{R}$, within which the motion of the gas is fully synchronized with the motion of the object. 

We have chosen representative orbital parameters of this scenario, i.e. the pericenter distance $R_{\rm \star,per}=50GM_{\rm BH}/c^2$, the inclination $i=45^{\circ}$ and the influence radius $\mathcal{R}=10GM_{\rm BH}/c^2$.
The 22-day eruptions period corresponds to a very long orbital period of $P_{\rm orb}=88000GM_{\rm BH}/c^3$, therefore we have simulated only one perturber passage through the pericenter in a 2D simulation. 
The resulting run is shown in Supplementary Figure 1, where we have chosen four different time instances illustrating the details of the transit. 
For each time instant, we show the density in a logarithmic scale and the corresponding map of the quantity called fast-outflowing rate density  $\dot{m}_{\rm out} = \rho \, u^r \sqrt{-g}$ for gas with Lorentz factor $\Gamma > 1.005$ and vanishing otherwise.
In the bottom panel, we show the time profile of the accretion rate and the fast-outflowing rate integrated through the upper and the lower funnel with the opening angle $45^{\circ}$ at a chosen diagnostic radius $R_{\rm diag}=100GM_{\rm BH}/c^2$.

The four time instances were chosen so that the main features of the transit are visible, in particular the strong density wave emerging directly during the passage ($t=10400GM_{\rm BH}/c^3 = 2.45d$); the launched big blob of matter which is immediately accelerated away along the funnel-disc boundary with a mildly relativistic speed, while the density perturbation has not yet propagated towards the horizon ($t=11000GM_{\rm BH}/c^3 = 2.59d$); the smaller blob pushed to the other direction, when the star is going back through the disc, while the accretion rate is still on the original level ($t=12100GM_{\rm BH}/c^3=2.85d$); and finally the abrupt rise of the accretion rate ($t=12200GM_{\rm BH}/c^3=2.87d$).

The main strength of the perturber-disc interaction model is that for an eccentric orbit with $0.73\lesssim e_{\star}\lesssim 0.97$ with the pericenter distance in the range $18.8\,R_{\rm g}\lesssim R_{\rm p}\lesssim 167.7\,R_{\rm g}$, which is sufficiently inclined with respect to the flow midplane ($\gtrsim 45^{\circ}$), the interaction can generate a sufficient accretion rate perturbation as well as launch a fast outflow with the velocity of $v\gtrsim 0.1c$, which may be relevant for some sources similar to \target. In addition, for the magnetically arrested disc, the perturber triggers reconnection events in a quasi-periodic way, which can temporally increase the inflow rate by several orders of magnitude as shown by GRMHD simulations of \cite{2021ApJ...917...43S}. The quasiperiodic behaviour of eruptions can be followed by a longer period of quiescence, hence the perturbed MAD flow can reproduce irregularities seen for the \target X-ray light curve. 

The main weakness of the model is that the soft X-ray spectrum with the black-body temperature of $\sim 100$--$200\,{\rm eV}$ is not achievable in the star-ADAF interaction regime due to the presence of high-temperature electron-proton plasma. 

Hence, an additional influx of stellar material at the pericenter due to the tidal Roche-lobe overflow is eventually required to provide a gas of lower temperature. Such a setup should eventually be studied in detail using GRMHD simulations with the appropriate heating/cooling terms included as well as the detailed treatment of shocks in stream-stream collisions.

{Recently, \cite{Linial2023} proposed an alternative accretion disk-perturber interaction model, where instead of driving a short-lived increase in the accretion rate, the perturber collides with the disk driving a shock and ejecting optically-thick gas clouds above and below the midplane, which produces the soft thermal X-ray emission. This model should be able to produce the correct temperature corresponding to the thermal X-ray emission, which is in contrast to the purely ADAF model. However, the lack of any evidence for the presence of a standard thin accretion disk in the \target's host prior and between the eruptions is hard to reconcile with the model. The same applies to other proposed interaction models involving thin disks \cite{2022arXiv221008023L}. In comparison with \cite{Linial2023}, who study star-disk collisions, \cite{2023A&A...675A.100F} prefer secondary black holes of $\sim 100\,M_{\odot}$ crossing the disk. Nevertheless, the thermal X-ray emission is produced similarly in adiabatically expanding, optically thick clouds pulled out from the accretion disk. The calculated eruption profiles of \cite{2023A&A...675A.100F} are inconsistent with \target since they exhibit a faster rise-slower decay. In a similar setup to \cite{Linial2023}, \cite{2023arXiv230403670T} studied star-disk crossings and the soft X-ray eruptions can be produced due to stellar bow-shock breakout emission. The main drawbacks of their model are fine-tuned parameters of the stellar orbit that needs to be low-inclination and retrograde in combination with the required short recurrence QPE timescale of $\lesssim 20$ hours, clearly inconsistent with \target. In addition, the radiatively efficient AGN disk is assumed in their model.

\subsubsection*{Stellar mass-transfer}
\label{subsubsec_stellar_mass_transfer}

A matter overflow from the star into the close vicinity of the MBH has been suggested to explain quasi-periodic X-ray eruptions. The overflow via an L1 point of the MBH-star binary system is expected to occur when the Roche-lobe of an orbiting star shrinks at its periapse so that a fraction of the stellar envelope is prone to tidal stripping. In the following estimates, we adopt the theoretical considerations analogous to e.g., \cite{2022ApJ...941...24K}, using the quantities inferred for \target X-ray eruptions: peak luminosity $L_{\rm peak}=6.0\times 10^{42}\,{\rm erg\,s^{-1}}$, inferred thermal temperature at the eruptions peak $k_{\rm B}T_{\rm peak}\sim 200\,{\rm eV}$ ($T_{\rm peak}\sim 2.3\times 10^6\,{\rm K}$), the mean eruption periodicity $P_{\rm eruption}\sim 22$ days, and the approximate duty cycle of the eruptions $D_{\rm eruption}\sim 7\,\text{days}/22\,{\rm days}\simeq 0.3$. 

Considering that X-ray spectra of the eruptions are consistent with being thermal, the eruption luminosity and the temperature constrain the geometric mean of the orthogonal length scales of an emitting area \cite{2022ApJ...941...24K},
\begin{align}
    R_{\rm emit}\simeq 2.45\times 10^{10}\left(\frac{L_{\rm peak}}{6\times 10^{42}\,{\rm erg\,s^{-1}}} \right)^{1/2} \left(\frac{T_{\rm peak}}{2.3\times 10^6\,{\rm K}} \right)^{-2}\,{\rm cm}\sim 0.35\,R_{\odot}\sim 0.04\,R_{\rm g}\,.
    \label{eq_thermal_radius}
\end{align}
which is comparable to the Solar radius and much smaller than the gravitational radius of $10^{6.6}\,M_{\odot}$ MBH. This strengthens the case for the association of the eruption with, e.g., the narrow circularization shock taking place close to the innermost stable circular orbit (ISCO), while the much smaller length-scale in comparison with the gravitational radius makes the accretion origin less likely.

As in the previous subsection, we consider a star on an eccentric orbit and assume that the X-ray eruption is triggered when the star approaches the pericenter of its orbit, hence $P_{\rm orb}\sim P_{\rm eruption}$, which again constrains the semi-major axis of the star in $R_{\rm g}$,
\begin{equation}
    a_{\star}\simeq 620.5 (P_{\rm orb}/22\,\text{days})^{2/3} (M_{\rm BH}/10^{6.6}\,M_{\odot})^{-2/3}\,R_{\rm g}.
\end{equation}

In comparison with other QPE sources whose period was of the order of hours and 
the duty cycle of 0.1 (see \cite{2022ApJ...941...24K,2022arXiv221008023L}), \target 
is in this regard a particular source with $P_{\rm eruption}\sim 22$ days and the 
duty cycle of $\sim 0.3$, i.e., the recurrence timescale is longer by a factor of 
$\sim 50$ and the duty cycle by a factor of $\sim 3$. This suggests different 
properties of a putative star in comparison with the stellar EMRI model of \cite{2022arXiv221008023L}.
To transfer the mass onto the MBH, the stellar pericenter distance needs to be comparable to the tidal radius, $R_{\rm p}\sim R_{\rm t}$, from which the stellar radius can be estimated as follows,
\begin{align}
    R_{\star}\simeq  \left(\frac{P_{\rm orb}^2 G m_{\star}}{4\pi^2} \right)^{1/3}\sim 33 \left(\frac{P_{\rm orb}}{22\,\text{days}} \right)^{2/3} \left(\frac{m_{\star}}{1\,M_{\odot}} \right)^{1/3}\,R_{\odot}\,,
    \label{eq_stellar_radius}
\end{align}
where the relation assumes an orbit with a small eccentricity of $e\ll 1$. Hence, a single star triggering the eruptions is expected to be significantly larger than a Solar-type star. From Eq.~\eqref{eq_stellar_radius} it is evident that for a given periodicity $P_{\rm orb}$, $R_{\star}\propto m_{\star}^{1/3}$, which is significantly flatter than the relation for main-sequence stars, $R_{\star}\propto m_{\star}^{0.8}$, see e.g. \cite{2023ApJ...945...86L}. For \target, both relations cross at $(m_{\star}/1\,M_{\odot})\sim 1798(P_{\rm orb}/22\,\text{days})^{10/7}$, which is an unrealistically large mass. Therefore, a single star orbiting the MBH must be an evolved star -- a red giant. A similar scenario where a star on an eccentric orbit evolves from the main sequence, fills its Roche lobe, and ``spoon feeds'' the SMBH at every pericenter was analyzed by \cite{2013ApJ...777..133M}, although for generally larger semi-major axes/periods.

To constrain the stellar orbit, we consider different orbital eccentricities and their relation to the Roche-lobe refilling timescale and the stellar mass loss. Using the Roche-lobe radius estimate at the given distance $r$ from the MBH, $R_{\rm RL}\sim r(m_{\star}/M_{\rm BH})^{1/3}$, we infer the effective stellar radius for which the duty cycle of the X-ray eruptions is 0.3. This is done by approximating the duty cycle as the ratio between the time when the stellar radius exceeds the corresponding Roche-lobe radius (and hence the stellar mass transfer can take place) and the total orbital period. The actual eruption duration may differ from this simplified estimate related to dynamics. However, we still expect a correlation between the Roche-lobe overflow timescale and the duty cycle. The non-zero eccentricity is clearly required to achieve the duty cycle less than unity. The emptied Roche lobe is expected to be refilled on the sound-crossing timescale $\tau_{\rm refill}\sim \Delta R_{\rm RL}/c_{\rm s}$. The sound speed for the stellar gas temperature $T\sim 10^6\,{\rm K}$ typical of stellar interior is $c_{\rm s}\sim 129\,{\rm km\,s^{-1}}$. To ensure sufficient time for refilling the Roche lobe for a new duty cycle, i.e., at most $(1-D_{\rm eruption})P_{\rm eruption}\sim 15.4$ days, $\tau_{\rm refill}<(1-D_{\rm eruption})P_{\rm eruption}$, which is met even for the case when the majority of the star is exposed to tidal stripping, i.e., for $e\sim 1$, $\tau_{\rm refill}<R_{\star}/c_{\rm s}\sim 2.1$ days. However, the formation of stellar EMRIs that are not tidally disrupted can be explained by the combined two-body relaxation$+$ gravitational-wave emission or the Hills mechanism$+$ gravitational-wave emission \cite{2022arXiv221008023L}, which limit the eccentricity, typically to values smaller than 0.5, though also an order of magnitude smaller semi-major axes. However, in the analysis by \cite{2022arXiv221008023L}, only main-sequence stars were considered, not evolved stars that also occupy nuclear star clusters. In the case of the Milky Way nuclear star cluster, late-type stars dominate the stellar composition (see \cite{2020A&A...641A.102S}), which provides further motivation to consider them as the potential candidates for longer-period stellar EMRIs.

To estimate the orbital eccentricity, we use the X-ray eruption peak luminosity of \target, $L_{\rm peak}\sim 6.0 \times 10^{42}\,{\rm erg\,s^{-1}}$ to calculate the necessarily detached stellar-mass per eruption whose free-fall energy is dissipated at the distance $R_{\rm diss}\sim 10\,R_{\rm g}$ close to the ISCO of the MBH,

\begin{align}
    \Delta M \sim 3.82 \times 10^{-5} \left(\frac{L_{\rm peak}}{6\times 10^{42}\,{\rm erg\,s^{-1}}} \right) \left(\frac{D_{\rm eruption}}{0.3} \right) \left(\frac{P_{\rm eruption}}{22\,\text{days}} \right) \left(\frac{R_{\rm diss}}{10\,R_{\rm g}} \right)   \,M_{\odot}\,,
\end{align}
while the corresponding mass-loss rate averaged over the orbital period is,
\begin{align}
    \dot{M}_{\star}\sim 6.34\times 10^{-4} \left(\frac{L_{\rm peak}}{6\times 10^{42}\,{\rm erg\,s^{-1}}} \right) \left(\frac{D_{\rm eruption}}{0.3} \right) \left(\frac{R_{\rm diss}}{10\,R_{\rm g}} \right)   \,M_{\odot}{\rm yr^{-1}}\,.
\end{align}
The stellar lifetime under the assumption of the constant mass-loss rate is $\tau_{\star}\sim m_{\star}/\dot{M}_{\star}\sim 1577 (m_{\star}/1\,M_{\odot}) (\dot{M}_{\star}/6.34\times 10^{-4}\,M_{\odot}{\rm yr^{-1}})^{-1}$ years, which is an upper limit due to unstable nature of the Roche-lobe overflow as we discuss further. 

The red giant with the radius of $R_{\star}\sim 33\,R_{\odot}$ and the mass $m_{\star}\sim 1\,M_{\odot}$ has a mean density of $\overline{\rho}_{\rm RG}\sim 3.9 \times 10^{-5}\,{\rm g\,cm^{-3}}$. Using a simple model of the mass removal between $R_{\star}$ and $R_{\rm RL}$ at the pericenter, we obtain $\Delta M\sim 3.8\times 10^{-5}\,M_{\odot}$ needed for one eruption for the Roche-lobe overfilling length-scale of $\Delta R_{\rm RL}=R_{\star}-R_{\rm RL}\sim 4.2 \times 10^{-4}\,R_{\odot}$, which corresponds to $e\sim 3.1 \times 10^{-5}$, i.e. a nearly circular orbit. For a smaller mean density, i.e. $\overline{\rho}_{\rm AGB}\sim 10^{-8}\,{\rm g\,cm^{-3}}$ typical of an AGB star with an initial radius of $R_{\star}\sim 500\,R_{\odot}$, the corresponding $\Delta M$ is reached for $\Delta R_{\rm RL}\sim 1.98\,R_{\odot}$ or $e\sim 0.12$. Depending on the stellar-envelope density, orbits with small eccentricities are sufficient to produce luminous eruptions as those for \target, which is consistent with the two-body relaxation in combination with the gravitational-wave emission that circularizes the orbit.

In addition to standard stellar evolution, one of the formation channels to produce inflated stars with $R_{\star}\sim 30\,R_{\odot}$ close to the MBH are stellar collisions. Considering a relaxed stellar cusp with the number density $n_{\star}\propto r^{-3/2}$ (Bahcall-Wolf cusp), the collisional timescale is $t_{\rm col}\propto (n_{\star} \sigma_{\star} R_{\star}^2)^{-1}$, where $\sigma_{\star}=(GM_{\rm BH}/a_{\star})^{1/2}$ is a local velocity dispersion. For a Solar-type star, the collisional timescale is $t_{\rm col}\sim 51\,000$ years, i.e. comparable to a TDE rate. The collisional stellar product stays heated and puffed up on the timescale comparable to or longer than $t_{\rm col}$, as is approximately given by the Kelvin-Helmholtz thermal timescale for an inflated star, $t_{\rm KH}=Gm_{\star}^2/(R_{\star}L_{\star})\sim 16\,000\,{\rm yr}$. Hence, the innermost stars can in principle be collisionally inflated and in appearance similar to the Galactic center G objects \cite{2019ApJ...878...58S}. On the other hand, once the star is puffed up to several tens of Solar radii, the collisional timescale decreases significantly since $t_{\rm col}\propto R_{\star}^{-2}$, i.e. for $R_{\star}\sim 30\,R_{\star}$, $t_{\rm col}\sim 56$ years, which can address the period irregularities, scarcity, and limited timescale of the stellar-transfer phenomenon in galactic nuclei.  

Since the distance of the star for the \target X-ray eruptions model is an order of magnitude further away than for the other QPE sources, the angular momentum of the detached stream needs to be effectively removed close to the pericenter. Magnetic stresses resulting from the turbulent dynamo \cite{2022ApJ...941...24K} or a prominent gas-drag force between the detached stellar stream and the ambient ADAF at the pericenter \cite{2014A&A...565A..17Z} are candidates for the efficient angular momentum removal. Under this assumption,the lower limit on the infall timescale is given by the free-fall timescale from the distance $a_{\star}$,
\begin{equation}
    \tau_{\rm ff}\sim \frac{\pi}{2\sqrt{2}} \left(\frac{a_{\star}}{R_{\rm g}} \right)^{3/2} \frac{GM_{\rm BH}}{c^3}\sim 3.9\,\left(\frac{a_{\star}}{620.5\,R_{\rm g}} \right)^{3/2} \left(\frac{M_{\rm BH}}{10^{6.6}\,M_{\odot}} \right)\,\text{days}\,,
\end{equation}
and hence $\tau_{\rm ff}< P_{\rm eruptions}$, which ensures that the X-ray eruptions can be discrete. 

In the model by \cite{2022ApJ...941...24K}, the X-ray emission is produced via stream-stream shocks due to apsidal precession close to the dissipation radius. The detached stellar material passes through L1 point and is subsequently tidally elongated and compressed, i.e., it undergoes spaghettification. The effective transverse size of the colliding streams $R_{\rm stream}$ needs to be sufficiently small to reach the eruptions peak temperature of $T_{\rm peak}\sim 2\times 10^6\,{\rm K}$,
\begin{align}
    T_{\rm peak}\sim 2.2\times 10^6 \left(\frac{L_{\rm peak}}{6\times 10^{42}\,{\rm erg\,s^{-1}}} \right)^{1/4} \left(\frac{R_{\rm stream}}{0.5\,R_{\odot}} \right)^{-1/2} \left(\frac{R_{\rm diss}}{10\,R_{\rm g}} \right)^{1/8}\,{\rm K},
    \label{eq_peak_temp}
\end{align}
which corresponds to $k_{\rm B}T_{\rm peak}\sim 190\,{\rm eV}$. The stream size $R_{\rm stream}$ in Eq.~\eqref{eq_peak_temp} is consistent with  the thermal length-scale limit given by Eq.~\eqref{eq_thermal_radius}. The X-ray eruptions can either be produced in shocks via stream-stream collisions as advocated by \cite{2022ApJ...941...24K} or by accretion. In the first case, the eruption duration is set by the timescale of the Roche-lobe overflow, which determines that the stream-stream collision can proceed for a sufficiently long time (cooling timescale of the shocked gas is much shorter than the eruption duration, $\tau_{\rm cool}\sim 240\,{\rm s}(L_{\rm peak}/6\times 10^{42}\,{\rm erg\,s^{-1}})$, hence the arrival time of the incoming stream needs to be of the order of the eruption duration). For the accretion scenario, a narrow radiating ring positioned at $R_{\rm diss}=10\,R_{\rm g}$ needs to be thicker than the standard disk. For the scale-height/radius ratio of $H/R\sim 0.1$ and the viscous parameter of $\alpha\sim 0.1$, the viscous timescale is $\tau_{\rm vis}\sim 7.2 (\alpha/0.1)^{-1}[(H/R)/0.1]^{-2}(R/10\,R_{\rm g})^{3/2}(M_{\rm BH}/10^{6.6}\,M_{\odot})$ days, which is comparable with the eruption duration. However, the accretion origin of X-ray eruptions is inconsistent with the eruption shape. For a ring that is accreted, the eruption rise should be shorter than its decline, while for \target the opposite is the case (see Extended Data Figure 2).

 The observed eruptions' periodicity, amplitude, as well as duration are variable. This is a general property of an unstable mass transfer \cite{2022arXiv221008023L,2022ApJ...941...24K}. As tidal forces peel off the upper stellar envelope, the star tends to puff up, which enhances the mass transfer initially. On the other hand, the stellar material inside the Roche lobe may also become temporarily depleted for the next few orbits, which can result in the ``eruption holidays". Stellar radial pulsations due to the kappa opacity mechanism typical of evolved supergiants (Cepheids) or low-mass stars (RR Lyrae) that are in the instability strip can also affect the amount of the transferred matter, and hence also the X-ray eruption detectability.

\subsubsection*{Interacting stellar EMRIs}

Apart from the single EMRI and the associated Roche-lobe mass overflow onto the MBH, the model involving two co-planar EMRIs on nearly circular orbits was suggested \cite{2022ApJ...926..101M}. The model consists of two stars with different masses, $m_1$ and $m_2$ with $m_1<m_2$, where the heavier star approaches the lighter one due to shorter gravitational-wave emission timescale.

In the further estimates, we set $m_1=1\,M_{\odot}$, $R_1=1\,R_{\odot}$ for the inner star and $m_2=1.5\,M_{\odot}$ and $R_2=1.4\,R_{\odot}$ for the outer star, where the radius was inferred using $R_{\star}\propto m_{\star}^{0.8}$ applicable for main-sequence stars. In principle, both stars can be undergoing Roche-lobe overflow. However, the mass transfer from the heavier outer star is significantly enhanced when the small separation of semimajor axes $\Delta a$ is reached. The semimajor axis of Roche-lobe overflowing stars must be comparable, $a_1\simeq a_2=a$,
\begin{equation}
    \frac{a}{R_{\rm g}}\simeq \frac{2.17c^2 R_{1,2}}{Gm_{1,2}^{1/3}M_{\rm BH}^{2/3}}\approx 40.8 \left(\frac{R_{1,2}}{1\,R_{\odot}} \right) \left( \frac{m_{1,2}}{1\,M_{\odot}}\right)^{-1/3} \left( \frac{M_{\rm BH}}{10^{6.6}\,M_{\odot}}\right)^{-2/3}\,. 
\end{equation}

The critical value for the orbital separation $\Delta a$ of both stars is (scaled to $R_2$),
\begin{align}
    \frac{\Delta a}{R_2}&\approx 5.8 \left(\frac{\chi}{\overline{\kappa}} \right)^{1/7} \left(\frac{M_{\rm BH}}{10^{6.6}\,M_{\odot}} \right)^{-4/21} \left(\frac{m_1}{1\,M_{\odot}} \right)^{9/14} \left(\frac{m_2}{1.5\,M_{\odot}} \right)^{-2/7} \left(\frac{R_2}{1.5\,R_{\odot}} \right)^{-3/28}\,\notag\\
    &\times \left(\frac{P_{\rm eruption}}{22\,\text{days}} \right)^{1/7} \left(\frac{T_{\rm eff}}{10^4\,{\rm K}} \right)^{-4/7}\,.
\end{align}
where the factor $\chi$ is of the order of unity, the stellar atmosphere opacity $\overline{\kappa}$ is scaled to electron scattering opacity, and $T_{\rm eff}$ is the atmosphere effective temperature. Given the orbital separation $\Delta a$, we can estimate the expected X-ray eruption recurrence timescale under the assumption $P_{\rm eruption}\sim T_{\rm fly}$. For the co-orbiting case, we obtain
\begin{align}
    P_{\rm eruption}\sim 17.4\left(\frac{M_{\rm BH}}{10^{6.6}\,M_{\odot}} \right)^{1/3}\left(\frac{m_{2}}{1.5\,M_{\odot}} \right)^{-5/6} \left(\frac{R_{2}}{1.4\,R_{\odot}} \right)^{3/2} \left(\frac{\Delta a}{5.8R_2} \right)^{-1}\,{\rm days},    
\end{align}
which is approximately consistent with the mean 22-day separation of X-ray eruptions in Swift J0230+28 given the involved factors of the order of unity.

The recurrence timescale of the X-ray eruption $P_{\rm eruption}$ can simply be associated with the pattern speed or a synodic period $P_{\rm s}$, i.e., the timescale on which stars approach each other. The required semimajor axis difference $\Delta a\sim 5R_2\sim 5R_{\odot}$ required for a significantly enhanced mass transfer implies the time difference $\Delta t=P_2-P_1$ between the orbital periods of stars 2 and 1, respectively. For the coorbiting case, we have $P_{\rm s}=(P_1^2+P_1\Delta t)/\Delta t$, while for the counter-orbiting case we obtain $P_{\rm s}=(P_1^2+P_1\Delta t)/(2P_1+\Delta t)$. Subsequently, these simple geometrical relations constrain the semimajor axis (and periods) for the two stars. For $P_{\rm eruption}\sim P_{\rm s}=22$ days, we obtain $a_1=44.9\,R_{\rm g}$ ($P_1=0.43$ days) and $a_2=45.5\,R_{\rm g}$ ($P_2=0.44$ days) for co-orbiting stars, which is close to the Roche-lobe overflowing limit for Solar-type, main-sequence stars. For counter-orbiting stars, the possible solution is $a_1=984.7\,R_{\rm g}$ ($P_1=43.98$ days) and $a_2=985.3\,R_{\rm g}$ ($P_2=44.02$ days), i.e., the eruptions would occur twice per orbital period. However, such a large semimajor axis would imply a large, atypical star in order for the Roche-lobe overflow to take place. Also, the coorbiting case is more plausible for two stars that have comparable semimajor axes due to gravitational wave emission since the Lense-Thirring differential nodal precession brings them to the same orbital plane on the precession timescale of $\sim 100$ days (see the estimate below) for distances of $\sim 40\,R_{\rm g}$ from the MBH. 

Under the assumption that the X-ray eruption is powered by the accretion of a narrow ring with the width of $\sim R_{\odot}$, see Eq.~\eqref{eq_thermal_radius}, its scale-height to radius ratio should be $h/r\sim 0.35$ to reproduce the eruption duty cycle of $D_{\rm eruption}\sim 0.3$,
\begin{align}
   D_{\rm eruption}\sim 0.3 \left(\frac{\alpha}{0.1} \right)^{-1} \left(\frac{h/r}{0.35} \right)^{-2} \left(\frac{m_1}{1\,M_{\odot}} \right)^{1/3} \left(\frac{M_{\rm BH}}{10^{6.6}} \right)^{-1/3} \left(\frac{R_{\rm circ}}{a} \right)^{3/2}\,,   
\end{align}
where $R_{\rm circ}$ represents the circularization radius of the flow. The duty cycle was estimated from the viscous timescale calculated for $R_{\rm circ}$. In case the circularization radius is smaller than $a$ due to the removal of the specific angular momentum of the gas by e.g. an outflow, then the formed disc can be even thinner: $h/r=0.12$ for $R_{\rm circ}\sim 10\,R_{\rm g}$. 

The expected mass transfer from the outer star onto the MBH during the critical separation $\Delta a$ (lasting about $10^4$ years) is,
\begin{align}
    \Delta M\sim 7\times 10^{-6} \chi^{-1}\left(\frac{M_{\rm BH}}{10^{6.6}\,M_{\odot}} \right) \left(\frac{m_{2}}{1.5\,M_{\odot}} \right)^{17/6} \left(\frac{R_2}{1.4\,R_{\odot}} \right)^{-7/2} \left(\frac{P_{\rm eruption}}{22\,\text{days}} \right) \left(\frac{\Delta a}{5.8 R_2} \right)^{-1}\,M_{\odot}\,, 
\end{align}
which leads to the bolometric luminosity of
\begin{equation}
L_{\rm eruption}\sim 2.0 \times 10^{42}\eta_{0.1} \left(\frac{\Delta M}{7\times 10^{-6}\,M_{\odot}}\right)\left(\frac{\tau_{\rm eruption}}{7\,\text{days}}\right)^{-1}\,{\rm erg\,s^{-1}}
\end{equation}
for the eruption duration of $\tau_{\rm eruption}\sim 7$ days and $10\%$ radiative efficiency. This is within the same order of magnitude as the X-ray eruptions of Swift J0230+28.

However, the compact accretion disc formed from the detached stellar material by circularization at $R_{\rm circ}\sim 10\,R_{\rm g}$ is expected to have a low effective temperature of $T_{\rm eff}\sim 75\,000\,{\rm K}$ (6.5 eV), which is inconsistent with the soft X-ray spectra during the eruptions implying a much higher temperature. The X-ray eruptions therefore appear to be powered by narrow stream-stream-collisions as discussed before for a single EMRI case (see Subsection `Stellar mass-transfer'}). The removal of the specific angular momentum of the stellar stream is easier for coorbiting interacting EMRIs since the Roche-lobe overflowing star is an order of magnitude closer to the innermost stable circular orbit.

The enhanced mass transfer powering quasiperiodic X-ray eruptions will cease or restart due to the nodal orbital precession, which is driven mainly by the Lense-Thirring differential precession. The nodal precession is present for any initially coplanar stellar orbits that are inclined by $\iota$ with respect to the equatorial plane of the MBH. The precession timescale can be estimated as,
\begin{equation}
    \tau_{\rm prec}\sim 103 \left(\frac{M_{\rm BH}}{10^{6.6}\,M_{\odot}} \right)^{-2} \left(\frac{\chi_{\rm BH}\sin{\iota}}{0.025} \right)\left(\frac{a}{40.8\,R_{\rm g}}\right)^3\,\text{days},
\end{equation}
where the factor $\chi_{\rm BH}\sin{\iota}$ is estimated for a small MBH spin of $\chi_{\rm BH}=0.05$ and the inclination of interacting stellar EMRIs of $\iota=30^{\circ}$ with respect to the MBH equatorial plane.

\subsubsection*{Compressed reformed clumps from a past TDE}

It has been proposed that the debris streams from TDEs are gravitationally unstable \cite{Coughlin2015} or thermo-dynamically unstable to Kelvin-Helmholtz (KH) interactions and line cooling \cite{Guillochon2014} and fragment into small-scale, localized knots. 
The instability becomes nonlinear (resulting in the knot formation) long after the disruption, because the shear along the stream inhibits the influence of self-gravity \cite{Coughlin2016, Coughlin2023}, or because the
KH instability growth rate requires the stream density to be comparable to the ambient density \cite{Chandrasekhar1961}. 
Therefore, 
the X-ray eruptions could be caused by the return of low-mass and low-density ``blobs'' of gas to the supermassive black hole (see \cite{Coughlin2020c} for a similar interpretation in the context of GSN069; \cite{Miniutti2019}) following a TDE, 
the TDE itself 
being undetectable because of, e.g., inefficient circularization (e.g., \cite{Guillochon2015}). 
The return of the blobs to pericenter is expected to be quasi-periodic 
if the blobs form out of a gravitational instability (see Equation 25 of \cite{Coughlin2020b}). 

If the clump formation process is approximately adiabatic and the stream remains self-gravitating (radiative recombination can raise the gas entropy, but not enough to destroy self-gravity; \cite{Coughlin2023}),  
the effective $\beta$ of each clump is much greater than one; here $\beta = R_{\rm t}/R_{\rm p}$, where $R_{\rm p}$ is the pericenter distance of the blob ($\sim$ equal to that of the original star)
and $R_{\rm t}$ is the tidal radius of the blob. If we let the mass of each clump be $M_{\rm clump} = x M_{\star}$ with $x \ll 1$, then we can show \cite{Nixon2022} 

\begin{equation}
    R_{\rm clump} = 
    R_{\star}x^{-1/3}, \,\,\, \rho_{\rm clump} = 
    \rho_{\star}x^2, \,\,\,  R_{\rm t, clump} = R_{\rm t}x^{-2/3}, \,\,\, \beta_{\rm clump} = \beta_{\star}x^{-2/3}. \label{Rclump}
\end{equation} 
Here $R_{\rm clump}$ is the clump radius, $\rho_{\rm clump}$  is the average clump density, $R_{\rm t, clump}$ is the clump tidal radius, and $\beta_{\rm clump}$ is the corresponding $\beta$ of the clump encounter with the black hole. The same variables but with subscript-$\star$ refer to the properties of the originally disrupted star. 

A plausible value for $x$ is $x \simeq 10^{-3}$ \cite{Coughlin2015}, implying $\beta_{\rm clump} \gtrsim 100$ for $\beta_{\star} \simeq 1$. For such high values of $\beta$, the clump is elongated in the direction of the MBH to a length \cite{Darbha2019} 

\begin{equation}
    L_{\rm clump} \simeq \frac{8}{5}R_{\rm clump}\beta_{\rm clump}^{1/2} \simeq \frac{8}{5}R_{\star}\beta_{\star}^{1/2}x^{-2/3}. \label{Lclump}
\end{equation}

and compressed within the plane 
by the factor $\simeq R_{\rm clump}\beta_{\rm clump}^{-1/2} \simeq R_{\star}$. 
Each clump is therefore transformed into a crescent of length $L_{\rm clump}$ that has an in-plane width comparable to the radius of the original star (e.g., Figure 1 of \cite{Bicknell1983}, or Figure 19 of \cite{Coughlin2022b}).

Each clump is also compressed by the vertical component of the tidal field of the MBH, reaching maximal compression near pericenter (e.g., \cite{Bicknell1983, Stone2013}). 
For these very high values of $\beta$, a weak shock forms during the compressive phase that propagates toward the orbital plane \cite{Coughlin2022b}, which then reflects off the midplane and propagates into the overlying and compressing material. Post-maximum-compression, the gas beneath the shock expands upward at a maximum velocity of 
\cite{Stone2013}

\begin{equation}
v_{\rm bounce} \simeq \beta_{\rm clump}\sqrt{\frac{2GM_{\rm clump}}{R_{\rm clump}}} \simeq \beta_{\star}\sqrt{\frac{2GM_{\star}}{R_{\star}}}, \label{vbounce}
\end{equation}
which is comparable to the escape speed of the original star if $\beta_{\star} \simeq 1$. As the gas rebounds, 
a contact discontinuity forms that separates the forward shock 
from a reverse shock,
all of which moves at a speed that is initially comparable to $v_{\rm bounce}$.

The temperature at the center of each reformed clump, still assuming that the entire process occurs adiabatically, is given by $T_{\rm clump} = T_{\star}x^{4/3}$. 

For $x = 10^{-3}$ and $T_{\star} \simeq 10^{7}$ K, this gives a temperature of $10^{3}$ K, but the photoionizing background of the galactic nucleus likely limits this to $\sim 10^4$ K \cite{Guillochon2014}. 
The shock heats the gas and raises the density to %
\cite{Coughlin2022b}

\begin{equation}
    T_{\rm max} \simeq 1.7T_{\rm clump}\beta_{\rm clump}^{1.12} \simeq 1.7 T_{\rm clump}\beta_{\star}^{1.12}x^{-0.75}, \,\,\,  \rho_{\rm max} = 1.6\rho_{\rm clump}\beta_{\rm clump}^{1.62} = 1.6\rho_{\star}\beta_{\star}^{1.62}x^{-0.92}. \label{Tmax}
\end{equation}
For $x = 10^{-3}$, $T_{\rm clump} = 10^4$ K, $\beta_{\star} = 1$, and $\rho_{\star} = 100$ g cm$^{-3}$, this gives $T_{\rm max} \simeq 3\times 10^{6}$ K and $\rho_{\rm max} \simeq 0.3$ g cm$^{-3}$. 
Equation \eqref{Tmax} shows that , 
each compressed clump reaches a temperature of $\simeq$ a few$\times 10^{6}$ K, and the density and temperature 
are large enough to render the gas optically thick.
As the gas rebounds, the forward shock is accelerated by the contact discontinuity 
until it reaches the edge of the collapsing material in front of it, at which point it accelerates 
\cite{Sakurai1960}. The nature of the acceleration depends on the properties of the envelope \cite{Ro13}, but it seems plausible that mildly relativistic velocities (characteristic of the UFO) could be reached. 

The gas temperature associated with \target is $\sim 2\times 10^{6}$ K and the peak luminosity is $L \simeq 5\times 10^{42}$ erg s$^{-1}$, 
and hence the size of the emitting region can be approximated as

\begin{equation}
    R_{\rm emit} \simeq \left(\frac{L}{\sigma T^{4}}\right)^{1/2} \simeq 5\times 10^{10} \textrm{ cm} \simeq 1R_{\odot}.
\end{equation}
This is consistent with the fact that the majority of the high-temperature emission is coming from the region of high compression, which is concentrated spatially near the pericenter of the returning clump. 
The duration of the eruption is set by the amount of time taken for the stretched clump to traverse the region of maximum compression; since this region occurs very near pericenter, this timescale is effectively the difference between the time it takes for the most-bound and least-bound fluid elements to reach pericenter, which is also the length of the clump divided by the velocity at pericenter. From Equation \eqref{Lclump} above, we therefore have

\begin{equation}
    t_{\rm X-ray} \simeq \frac{L_{\rm clump}}{v_{\rm t}} \simeq \frac{8}{5}\frac{R_{\rm clump}\beta_{\rm clump}^{1/2}}{\sqrt{\frac{GM}{R_{\rm p}}}} \simeq \frac{8}{5}\frac{R_{\star}\beta_{\star}^{1/2}}{\sqrt{\frac{GM}{R_{\rm p}}}}x^{-2/3}. \label{txray}
\end{equation}
Putting into this expression the canonical values $R_{\star} = 1R_{\odot}$, $\beta_{\star} = 1$, $R_{\rm p} = R_{\rm t} = R_{\star}\left(M_{\rm BH}/M_{\star}\right)^{1/3}$, $M = 10^6M_{\odot}$, and $x = 10^{-3}$ gives $t_{\rm X-ray} \simeq 1$ hr, which is close to the duration of the eruptions from some QPEs (e.g., Extended Data Figure 8) and perhaps some of the very short-duration eruptions from \target, but is too short to explain the $\sim$ few-day-long eruptions. There are a few possibilities that can reconcile this discrepancy: If 1) The originally disrupted star was a massive star with mass (e.g.) $M_{\star} = 20M_{\odot}$ and $R_{\odot} = 20R_{\odot}$, then Equation \eqref{txray} gives $t_{\rm X-ray} \simeq 1.6$ days; 2) The pericenter distance of the original star was larger than the canonical tidal radius (i.e., it was a partial disruption; this would be consistent with the lack of detection of the original TDE due to inefficient circularization, as the relativistic advance of periapsis angle declines linearly with pericenter distance); 3) The long-duration eruptions are actually many smaller eruptions temporally blended together; 4) The thermodynamic evolution of these clumps is not well-approximated as adiabatic, then the clump size could conceivably be much larger than the estimate given in Equation \eqref{Rclump}, especially if the equation of state is close to isothermal; 5) The clump mass is much smaller than our fiducial choice of $x = 10^{-3}$; 6) The pericenter distance of the returning clump is larger than the pericenter distance of the original star, which could occur if there are significant dynamical interactions between the stream material and a Keplerian flow (as presumably the latter has more angular momentum than the former).

\renewcommand{\figurename}{Supplementary Information Figure}
\renewcommand{\tablename}{Supplementary Information Table}
\setcounter{figure}{0}
\setcounter{table}{0}

\clearpage
\begin{table}[h!]
    
    \centering
    \ttabbox[\linewidth]{
   
    \begin{tabular}{ccccccc}

    \textbf{Epoch}    &   \textbf{Instrument} &  \textbf{Peak X-ray flux\footnote{In the 0.3-0.8 keV band for \nicer/XTI, and in 0.3-2.0 keV band for \swift/XRT}}   & \textbf{MJD Peak}    & \textbf{$\sigma_{+}$} & \textbf{$\sigma_{-}$} & \textbf{FWHM} \\
         &      & [$\times 10^{-13}$ \ergs]     &     &  [days]  &  [days] & [days]         \\
    \hline
 \textit{E3} & \swift &$17.6 \pm 1.6$ & $59797.4 \pm 0.4$ & $1.41 \pm 0.20$ & $2.4 \pm 0.4$ & $4.5 \pm 0.5$ \\
 \textit{E4} & \nicer &$21.3 \pm 0.4$ & $59821.1 \pm 0.2$ & $2.43 \pm 0.14$ & $2.9 \pm 0.1$ & $6.2 \pm 0.2$ \\
 \textit{E5} & \nicer &$16.6 \pm 0.4$ & $59845.4 \pm 0.1$ & $0.79 \pm 0.03$ & $1.3 \pm 0.1$ & $2.5 \pm 0.1$ \\
\textit{E6} & \swift & $11.1 \pm 1.1$ & $59858.7 \pm 0.3$ & $2.03 \pm 0.32$ & $3.1 \pm 0.3$ & $6.0 \pm 0.5$ \\
\textit{E10} & \swift & $18.3 \pm 1.7$ & $59951.5 \pm 0.2$ & $1.63 \pm 0.13$ & $1.9 \pm 0.1$ & $4.1 \pm 0.2$ \\
\textit{E11} & \swift & $13.2 \pm 1.7$ & $59969.5 \pm 0.3$ & $1.62 \pm 0.39$ & $2.3 \pm 0.3$ & $4.6 \pm 0.6$ \\
    \hline
    \end{tabular}
    }
    {\caption{{\bf Best-fitted parameters for the eruption shape fitting.}}
     \label{tab:eruption_shape} }

\end{table}

\begin{figure*}[h!]
    \centering
    \includegraphics[width=0.6\textwidth]{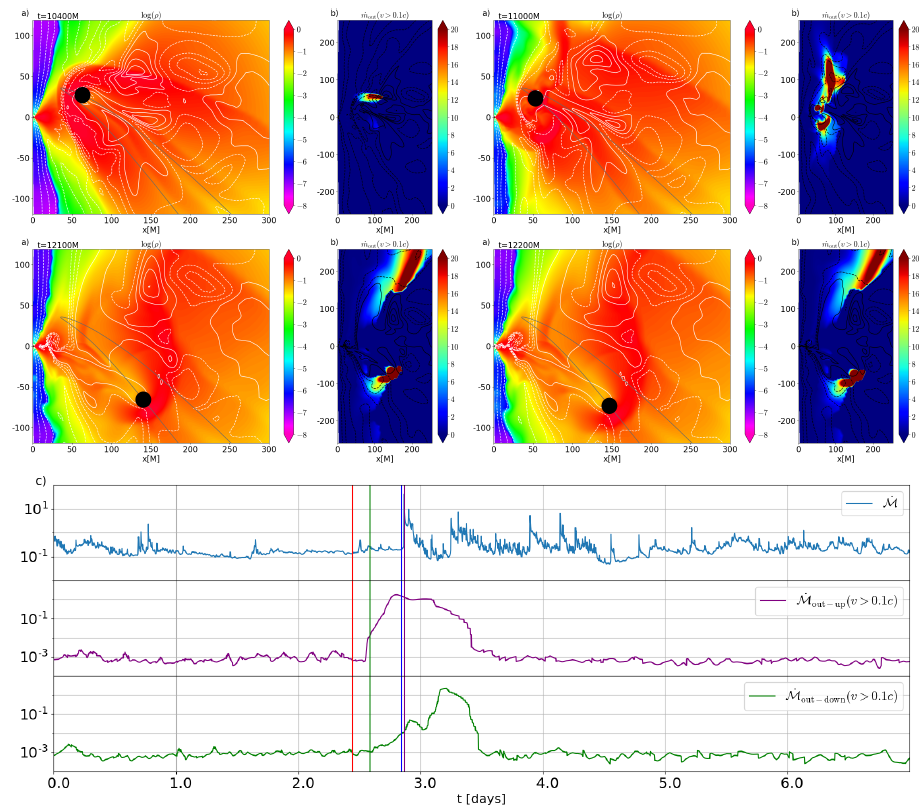}
    
    \caption{\textbf{Slices from GRMHD simulation of Accretion disc -- perturber interaction model}. The influence radius of the perturber $\mathcal{R}=10GM_{\rm BH}/c^2$, the SMBH mass $M_{\rm BH}=10^{6.6}M_\odot$. {\bf Top}: Panels with density in logarithmic scale and the fast-outflowing rate at four different time instances (all in dimensionless units).  {\bf Bottom:} The accretion rate and fast-outflowing rate ($v>0.1c$) in the upper half-funnel and lower half-funnel. The time instances of the slices are shown by vertical lines, $t=10400 GM_{\rm BH}/c^3 = 2.45 {\rm d}$ (red), $t=11000 GM_{\rm BH}/c^3 = 2.59 {\rm d}$ (green), $t=12100 GM_{\rm BH}/c^3 = 2.85 {\rm d}$ (blue), $t=12200 GM_{\rm BH}/c^3 = 2.87 {\rm d}$ (purple).}
    \label{fig:GRMHD_slice}
\end{figure*}

\clearpage
\begin{table}[ht!]
    \centering
    \ttabbox[\linewidth]{
    \begin{tabular}{cccc}
    \label{tab:L_x_T}
    {Start}    &   {End}   & {${\rm log}\  L_{\rm {0.3-0.8 keV}}$}          & {$T_{in}$} \\
       (MJD)  & (MJD)      & [{\rm \ergs}]          &  [eV]          \\
    \hline
    59798.0   & 59800.0   & 41.97 $\pm$ 0.02 & 153$^{+6}_{-5}$   \\
    59802.4 & 59802.5 & 42.78 $\pm$ 0.01 & 196$^{+8}_{-7}$   \\
    59814.0   & 59816.0   & 41.91 $\pm$ 0.03 & 113$^{+7}_{-7}$   \\
    59816.0   & 59817.0   & 42.13 $\pm$ 0.02 & 110$^{+4}_{-4}$   \\
    59817.0   & 59818.0   & 42.23 $\pm$ 0.02 & 123$^{+4}_{-4}$   \\
    59818.0   & 59820.0   & 42.51 $\pm$ 0.02 & 137$^{+5}_{-5}$   \\
    59820.0   & 59822.0   & 42.58 $\pm$ 0.01 & 172$^{+3}_{-3}$   \\
    59822.0   & 59824.0   & 42.46 $\pm$ 0.01 & 189$^{+6}_{-5}$   \\
    59824.0   & 59824.7 & 42.36 $\pm$ 0.01 & 174$^{+5}_{-5}$   \\
    59825.5 & 59825.7 & 42.31 $\pm$ 0.03 & 225$^{+18}_{-15}$ \\
    59841.0   & 59842.0   & 41.74 $\pm$ 0.01 & 115$^{+3}_{-3}$   \\
    59842.0  & 59843.0   & 41.88 $\pm$ 0.01 & 115$^{+3}_{-3}$   \\
    59843.0   & 59844.0   & 42.08 $\pm$ 0.02 & 113$^{+4}_{-4}$   \\
    59844.0   & 59845.0   & 42.41 $\pm$ 0.01 & 158$^{+3}_{-3}$   \\
    59845.0   & 59846.0   & 42.46 $\pm$ 0.01 & 185$^{+4}_{-4}$   \\
    59846.0   & 59846.6 & 42.30 $\pm$ 0.01 & 185$^{+5}_{-5}$   \\
    59850.0   & 59852.0   & 41.92 $\pm$ 0.01 & 108$^{+3}_{-3}$   \\
    59852.0   & 59854.0   & 42.16 $\pm$ 0.01 & 131$^{+4}_{-3}$   \\
    59854.0   & 59857.9 & 42.37 $\pm$ 0.01 & 136$^{+3}_{-3}$   \\
    59857.9 & 59860   & 42.48 $\pm$ 0.01 & 182$^{+8}_{-7}$   \\
    59878.8 & 59879.2 & 42.13 $\pm$ 0.01 & 136$^{+3}_{-3}$   \\
    59882.0   & 59882.4 & 42.14 $\pm$ 0.01 & 137$^{+4}_{-4}$   \\
    59949   & 59951.2 & 42.75 $\pm$ 0.01 & 164$^{+5}_{-5}$   \\
    59951.4 & 59953.5& 42.66 $\pm$ 0.01 & 218$^{+8}_{-7}$   \\
    \hline
    \end{tabular}
    }
    {\caption{{\bf Summary of time-resolved X-ray spectra analyses with pure thermal model}. 
    Here,  0.3--0.8 keV \nicer~ spectra are fit with {\tt tbabs*zashift*diskbb} model using 
    {\it XSPEC} \cite{xspec}. {\bf Start} and {\bf End} represent the start and end times (in 
    units of MJD) of the interval used to extract a combined \nicer spectrum. {\bf $\log 
    L_{\rm 0.3-0.8 keV}$} is the logarithm of the integrated absorption-corrected luminosity 
    in 0.3--0.8 keV in units of \ergs. {\bf $T_{in}$} is the best-fitting inner disk 
    temperature of \texttt{diskbb} in eV.  Uncertainties represent 1$\sigma$ level.}}

\end{table}

\end{document}